\documentclass[11pt]{article}

\usepackage[includeheadfoot,bindingoffset=0.5in,inner=0.5in,outer=0.75in,top=0.5in,bottom=0.5in]{geometry}
\usepackage{enumerate}
\usepackage{amssymb,amsmath,amsfonts}
\usepackage{comment}
\usepackage{color}
\usepackage[T1]{fontenc}
\usepackage[utf8]{inputenc}
\usepackage{authblk}
\numberwithin{equation}{section}
\usepackage{graphicx}
\graphicspath{{.}}
\usepackage[font=small,labelfont=bf,tableposition=top]{caption}
\usepackage[font=footnotesize]{subcaption}
\usepackage{fixltx2e}

\newcommand{\U}{\mathbf{U}}
\newcommand{\FH}{\mathbf{F}_H}
\newcommand{\FK}{\mathbf{F}_K}
\newcommand{\FD}{\mathbf{F}_D}
\newcommand{\FS}{\mathbf{F}_S}

\newcommand{\W}{\mathbf{W}}
\newcommand{\Dftoc}{\mathbf{D}^{f \rightarrow c}}
\newcommand{\Dntoc}{\mathbf{D}^{n \rightarrow c}}
\newcommand{\Dcton}{\mathbf{D}^{c \rightarrow n}}
\newcommand{\Gctof}{\mathbf{G}^{c \rightarrow f}}
\newcommand{\Gcton}{\mathbf{G}^{c \rightarrow n}}
\newcommand{\Gntoc}{\mathbf{G}^{n \rightarrow c}}

\newcommand{\IdentityTensor}{\mathbb{I}}
\newcommand{\HeatFlux}{\boldsymbol{\mathcal{Q}}}
\newcommand{\StressTensor}{\boldsymbol{\tau}}
\newcommand{\ReversibleStress}{\boldsymbol{\Pi}}
\newcommand{\KortewegStress}{\boldsymbol{\Pi}^\mathrm{K}}
\newcommand{\KortewegCross}{\boldsymbol{\Pi}^\mathrm{C}}
\newcommand{\ShearViscosity}{\eta}
\newcommand{\BulkViscosity}{\zeta}
\newcommand{\ThermalConductivity}{\lambda}
\newcommand{\half}{\frac{1}{2}}

\title{Modeling Multi-phase Flow using Fluctuating Hydrodynamics}
\date{}
\author[1]{Anuj Chaudhri \thanks{anuj.chaudhri@gmail.com,achaudhri@lbl.gov}}
\author[1]{John B. Bell}
\author[2]{Alejandro L. Garcia}
\author[3]{Aleksandar Donev}
\affil[1]{Computational Research Division, Lawrence Berkeley National Lab, Berkeley, CA 94720, USA}
\affil[2]{Department of Physics and Astronomy, San Jose State University, San Jose, CA 95192, USA}
\affil[3]{Courant Institute of Mathematical Sciences, New York University, New York, NY 10012, USA}

\begin{document}

\maketitle
\renewcommand{\abstractname}{\vspace{-\baselineskip}}
\begin{abstract}

Fluctuating hydrodynamics provides a model for fluids at mesoscopic scales where thermal fluctuations can have a significant impact on the behavior of the system.  Here we investigate a model for fluctuating hydrodynamics of a single component, multiphase flow in the neighborhood of the critical point.  The system is modeled using a compressible flow formulation with a van der Waals equation of state, incorporating a Korteweg stress term to treat interfacial tension. We present a numerical algorithm for modeling this system based on an extension of algorithms developed for fluctuating hydrodynamics for ideal fluids. The scheme is validated by comparison of measured structure factors and capillary wave spectra with equilibrium theory. We also present several non-equilibrium examples to illustrate the capability of the algorithm to model multi-phase fluid phenomena in a neighborhood of the critical point. These examples include a study of the impact of fluctuations on the spinodal decomposition following a rapid quench, as well as the piston effect in a cavity with supercooled walls. The conclusion in both cases is that thermal fluctuations affect the size and growth of the domains in off-critical quenches.

\end{abstract}

\section{Introduction}

\indent

At a molecular scale, the state of a fluid, as measured by hydrodynamic variables, is constantly changing and is stochastic in nature. Micro / nano fluidic devices operate in regimes in which these thermal fluctuations play an important role in the overall dynamics but where molecular level simulation is often computationally infeasible. Consequently, it is becoming increasingly important to develop models that can incorporate the effect of fluctuations in the continuum equations. The effect of fluctuations is important in both equilibrium \cite{Pathria_2007} and non-equilibrium \cite{deZarate_2006} systems. They also become important in complex chemical and biological systems where multiple time and length scales are involved, such as in a colloidal suspension where time and length scales can span over 15 orders of magnitude \cite{Padding_2006}. The effect of fluctuations have been shown to be important in the breakup of nanojets \cite{Moseler_2000,Eggers_2002,Kang_2007}, Brownian molecular motors \cite{Astumian_2002,Oster_2002,Vandenbroeck_2004,Meurs_2004}, Rayleigh-B\'{e}nard convection \cite{Wu_1995,Quentin_1995}, Kolmogorov flow \cite{Bena_1999,Bena_2000,Mansour_1999}, Rayleigh-Taylor mixing \cite{Kadau_2004,Kadau_2007}, combustion and explosive detonation \cite{Nowakowski_2003,Lemarchand_2004}, reaction fronts \cite{Moro_2004}, capillary dynamics \cite{Buscalioni_2008,Shang_2011}, thin films \cite{Mecke_2005,Grun_2006,Fetzer_2007}, drop formation \cite{Hennequin_2006}, fluid mixing \cite{Vailati_2011,Donev_2011} and multispecies mixtures \cite{Balakrishnan2014}. Fluctuations also need to be incorporated in advanced multiscale methodologies such as hybrid multiscale methods \cite{Donev_2010_1}, where the use of deterministic equations in the continuum leads to errors in the fluctuation spectrum of a coupled particle simulation.

Thermal fluctuations were first incorporated into the deterministic Navier-Stokes equations by Landau and Lifshitz \cite{Landau_1959}. The central idea in fluctuating hydrodynamics is to treat the dissipative fluxes as stochastic variables and incorporate an additional stochastic flux into the deterministic Navier-Stokes equations. These fluxes are the macroscopic manifestation of microscopic degrees of freedom in a thermodynamic system. These microscopic degrees of freedom, which are not directly modeled, give rise to hydrodynamic fluctuations and cause Brownian motion. This procedure is called \textit{coarse-graining} and can be done formally using Projection Operator techniques \cite{Espanol_2004,Zwanzig_1960,Grabert_1982,Ottinger_2005} that give rise to transport equations for the 'relevant' (slow) conserved quantities. The 'irrelevant' (fast) variables are eliminated but affect the conserved variables and cause them to fluctuate about their mean values. The justification for the linearized hydrodynamic fluctuations using Landau-Lifshitz, fluctuating Navier-Stokes (FNS) was given initially by Fox \cite{Fox_1970,Fox_1978} and independently by Bixon and Zwanzig \cite{Bixon_1969}. The non-linear hydrodynamic fluctuations have been justified based on derivation of the Fokker-Planck equation for the distribution function of coarse-grained densities of conserved quantities \cite{Zubarev_1983,DiscreteLLNS_Espanol} or the closely-related stochastic differential equations for the random fields \cite{Espanol_1998}.

Multiphase flows, such as gas/liquid and liquid/liquid mixtures, are not only of fundamental interest but also extremely important in many engineering applications (e.g., refrigeration, petroleum and chemical processing, electronics cooling, power plants, spacecraft technology). The phase change processes that occur are very complex since they involve the interplay of multiple length and time scales, thermal fluctuations, nonequilibrium effects and other effects due to dynamic interfaces. Hence developing new methodologies that are capable of capturing fundamental phase change behavior along with the complex fluid flow and heat transfer in the system are extremely important. Additional challenges arise in the form of developing numerical methods to solve these complex problems.  Over the last few years, much progress has been made in developing numerical schemes for fluctuating hydrodynamics in single phase fluids \cite{Donev_2010_2,Usabiaga_2012,Delong_2013,LowMachExplicit}. But multiphase systems offer newer challenges not only due to variation of properties from one phase to another but also due to the delicate fluctuation-dissipation balance that has to be maintained in the numerical schemes \cite{Atzberger_2007,Donev_2010_2}. 

The hydrodynamic equations for a general multicomponent, multiphase system have been derived systematically from the underlying microscopic dynamics within the GENERIC framework \cite{Espanol2003}. This model was used to study diffusion in binary fluid mixtures using a discrete model formulated in a smoothed particle hydrodynamics framework \cite{Thieulot2005smoothed1,Thieulot2005smoothed2,Thieulot2005non}. Mesoscale multiphase simulations have also been performed using Lattice-Boltzmann (LBM) techniques (see \cite{Aidun_2010_LBM} and references therein). While most of the models have been limited to incompressible, isothermal systems without fluctuations, significant progress is being made to extend this to non-isothermal situations \cite{Meng_2013_LBM} and incorporate thermal fluctuations \cite{Adhikari_2005,Dunweg_2007,Dunweg_2008,Gross_2010,Thampi_2011,Gross_2012}. Previous work on simulating multiphase systems using fluctuating hydrodynamics has been limited to isothermal systems \cite{Shang_2011}. In particular, Shang et al. \cite{Shang_2011} used the fluctuating hydrodynamics methodology along with the Ginzburg-Landau free energy to model liquid-vapor and liquid-liquid interfaces. They mapped fluctuations in density and height of the interface from molecular dynamics simulations (MD) to the hydrodynamic field variables and showed that the isothermal compressibility, surface tension and capillary wave spectrum are well reproduced by the fluctuating hydrodynamics equations. The work here represents an extension of their work to non-isothermal and non-equilibrium systems. 

The methodology described here is based on a diffuse interface model with an emphasis on capturing the correct surface tension between the two phases and serves as an extension of the fluctuating Navier Stokes (FNS) equations to a multiphase setting, based on the single-phase algorithm discussed in \cite{Balakrishnan2014}. Diffuse interface models have been developed and analyzed for multiphase systems since the time of van der Waals \cite{Rowlinson_2002,Anderson_1998}, for both single component and binary fluids. For binary fluids, there is an extensive body of literature \cite{Anderson_1998,Kim_2012} starting with the work of Cahn and Hilliard \cite{Cahn_1958,Cahn_1959_1,Cahn_1959_2}. For single-component, multiphase systems, most of the analysis has focused on using renormalization group techniques to look at critical point scaling laws for Model H, which describes the dynamics of a fluid phase transition as well as a single-component fluid near its critical point \cite{Hohenberg_1977}. Numerical models have focused on either the deterministic Navier-Stokes equations for isothermal van der Waals fluids \cite{Nadiga_1996,Lamorgese_2009,Pecenko_2010} or modified thermodynamic behavior of fluids with artificially thickened interfaces \cite{Jamet_2001}. To model the transport of energy or heat, Onuki introduced a dynamic model that allows temperature gradients to exist below the critical point or in a nonequilibrium setting that accounts for the correct latent heat transport \cite{Onuki_2007}. This model has been developed further to analyze boiling in van der Waals fluids \cite{Laurila_2012}.

In this paper we present a stochastic method of lines discretization for the compressible fluctuating hydrodynamics equations with a van der Waals equation of state. In Section II we present the theory for the diffuse interface method that is used to model the order parameter (density) as a smooth variation across the interface. The surface tension effects give rise to Korteweg type stresses that appear as additional terms in the momentum and energy equations. Section III presents the finite volume method used for the spatial discretization and a three stage explicit Runge-Kutta scheme used for temporal integration. The selection of a good value of the gradient energy coefficient ($\kappa$) is first discussed in section IV followed by validation of the numerical scheme by comparing the 2D and 3D structure factors with theory and capturing the correct capillary wave spectrum in fluctuating planar liquid-vapor interfaces. Two numerical examples are selected to demonstrate the capabilities of the model in Section V. In the first case, spinodal decomposition is modeled in a near-critical Argon system. Calculations of the structure factor for density show that the growth of droplets (in off-critical quenches) is enhanced when thermal fluctuations are included, whereas no such conclusions can be drawn for bubbles or for critical quenches (bicontinuous pattern). The second example looks at adiabatic cooling in a square cavity, induced by the piston effect~\cite{PistonEffect1,PistonEffect2}, when the boundary temperature is lowered to sub-critical temperatures. In this example, we also find that thermal fluctuations enhance droplet growth. The paper concludes with a summary and a discussion of future work.

\section{Fluctuating Hydrodynamics for a van der Waals Fluid}

For a single component fluid, the order parameter is chosen to be the average mass density of the fluid. This order parameter is modeled by the continuity equation itself and does not warrant an additional equation, such as in the Cahn-Hilliard model, though Shang et al. \cite{Shang_2011} have developed a more general set of equations where both the density and an order parameter can vary. This section first describes the free energy model based on the van der Waals equation of state followed by a description of the thermodynamics using the GENERIC framework \cite{Ottinger_2005}. The FNS equations are then described for the complete multiphase model.

\subsection{van der Waals Free Energy Model}

To model an inhomogeneous fluid such as a liquid-vapor system with a finite surface tension and interfacial width, it is first necessary to establish the statistical mechanical basis of such a model \cite{Rowlinson_2002,Evans_1979,Evans_1992}. An inhomogeneous fluid is best understood by the theory of density functional methods that are based on the idea of expressing the free energy of the inhomogeneous fluid as a functional of the mass density $\rho\left(\textbf{r}\right)$. Once this free energy functional is known, all relevant thermodynamic properties of the system can be calculated and in particular the surface tension can be computed for the interface. The Helmholtz free energy, $F$, (or equivalently the grand potential) can be expressed in terms of density, its gradient, and temperature $T$ \cite{Callen_2006}. Using symmetries and truncating the gradient expansion at the gradient square term,
\begin{equation}
\label{Helmholtz_Free_Energy_Eqn}
F\left(\rho(\textbf{r}),\nabla \rho(\textbf{r}) ,T(\textbf{r})\right) = \int \mathrm{d}\textbf{r} \, \left(f \left(\rho(\textbf{r}),T(\textbf{r})\right)
+ \half \kappa\left(\rho(\textbf{r}),T(\textbf{r})\right)\left|\nabla \rho(\textbf{r}) \right|^{2}\right)
\end{equation}
\indent The first term $f \left(\rho(\textbf{r}),T(\textbf{r})\right)$ contains all the local contributions to the free energy, which includes the hard-core repulsions as well as the short-range attractions. The square gradient term models the interfacial energy that is necessary to have liquid-vapor separation. Although written in continuum notation, the physical meaning of (2.1) in the theory of coarse-graining is of a bare free energy function that corresponds to coarse-graining the particles into cells of size sufficiently large to contain many particles (and thus justify a local free energy functional and separation of time scales). Importantly, the cell size needs to also be substantially smaller than the long-range tail of the intermolecular potential so that the attractive part of the internal potential can be written as an integral. Lastly, the variation of the density needs to be slow so that a Taylor series approximation can be made to obtain the square gradient term, which can be justified for slowly varying density profiles that occur close to the liquid-vapor critical point. 

\indent The square gradient term is also a useful approximation for modeling fluid-fluid interfaces even far from the critical point, as has been demonstrated through direct comparisons with molecular dynamics data \cite{Shang_2011}. The gradient energy coefficient, $\kappa\left(\rho(\textbf{r}),T(\textbf{r})\right)$, models the strength of the interfacial free energy and can be directly related to the surface tension. In general this gradient energy coefficient is a function of density and temperature; taking it as a constant reduces the above formalism to the van der Waals square gradient model. The van der Waals model can be justified by considering a liquid in which the intermolecular potential has a short-ranged repulsive core (excluded volume), as well as a weak but long-ranged attractive tail that is responsible for the square-gradient term in the free-energy \cite{Kampen_1964,EspanolVdW}.

The GENERIC framework provides a framework for non-equilibrium thermodynamics of multiphase systems \cite{Ottinger_2005}. In this two-generator framework, one specifies the total energy and the entropy as functionals of the conserved hydrodynamic fields. For a system described by the local mass, momentum and internal energy density fields $\rho(\textbf{r}),\textbf{g}(\textbf{r}),u(\textbf{r})$, the total energy $U$ and entropy $S$ can be defined by \cite{EspanolVdW},

\begin{equation}
\label{TotalEnergy_Eqn}
U = \int \mathrm{d}\textbf{r} \,  \left(\dfrac{1}{2}\dfrac{\left|\textbf{g}(\textbf{r})\right|^{2}}{\rho(\textbf{r})} + u (\textbf{r}) + \dfrac{1}{2}\kappa \left|\nabla \rho(\textbf{r}) \right|^{2} \right) 
\end{equation}

\begin{equation}
\label{TotalEntropy_Eqn}
S = \int \mathrm{d}\textbf{r} \, s \left(\rho(\textbf{r}),u(\textbf{r})-u^{\textrm{att}}(\textbf{r})\right) 
\end{equation}

\noindent where $\kappa$ is assumed to be a constant, $u = \rho e$ where $e$ is the specific internal energy, and $s$ is the local entropy density of a system at thermodynamic equilibrium at a given mass and internal energy density. It is important to note that $u(\textbf{r})$ is the local internal energy density, which contains interactions from the hard core potential as well as the short-range attractive interactions $u^{\textrm{att}}(\textbf{r})$ (=$-a^{\prime} n^{2}$ given in Eq.(\ref{VdW_Energy_Density_Eqn}) below). The entropy on the other hand is defined by the 'intrinsic' internal energy $u (\textbf{r})-u^{\textrm{att}}(\textbf{r})$, which contains interactions only from the hard core potential \cite{EspanolVdW}. The local free-energy density $f = u - Ts$ contains interactions from hard core as well as the attractive potential. Note that it is possible to also include square gradient terms in the entropy density \cite{Jelic_2009}; here we include only the dominant energetic contribution of the inhomogeneity of the fluid \cite{EspanolVdW}. For a thermodynamically admissible model using the above definitions for total energy and entropy, the single component multiphase model can be formulated within the GENERIC framework \cite{Grmela_Ottinger_1997,Ottinger_Grmela_1997,Ottinger_2005} and the appropriate equations can be written down \cite{EspanolVdW,Jelic_2009}, as given in the next section.

For the system to exhibit phase separation the equation of state should have a van der Waals loop \cite{Goldenfeld_1992}. The standard model is the van der Waals equation of state,
\begin{equation}
\label{VdW_EoS_Eqn}
P\left(\rho, T\right) = \dfrac{n k_{B}T}{1-b^{\prime} n} - a^{\prime} n^{2}
\end{equation}
\noindent where $P$ is the pressure, $k_{B}$ is the Boltzmann constant, and $n = \rho/m$ is the number density, with $m$ being the molecular mass. The van der Waals parameters, $a^{\prime}$ and $b^{\prime}$, indicate the strength of the long-ranged attractive forces and the excluded volume due to the short-ranged repulsive forces, respectively. From this equation of state, the number density, pressure, and temperature at the critical point are,
\begin{equation}
\label{VdW_Critical_Eqn}
n_{c} = \dfrac{1}{3b^{\prime}}, \qquad
P_{c} = \dfrac{a^{\prime}}{27 (b^{\prime})^{2}}, \qquad
T_{c} = \dfrac{8 a^{\prime}}{27 b^{\prime} k_{B}}
\end{equation}
\noindent so that $P_c / n_c k_B T_c = 3/8$.

Using basic thermodynamic definitions and some simple manipulations, the free energy density $f$ and the internal energy density, $u$, can be written as,
\begin{equation}
\label{VdW_Helmholtz_Eqn}
f = n k_{B}T \ln \left[ \dfrac{\rho}{1-b^{\prime} n} \right] - a^{\prime}n^{2}
\end{equation}
\noindent and
\begin{equation}
\label{VdW_Energy_Density_Eqn}
u = \dfrac{3}{2} n k_{B}T - a^{\prime} n^{2} \;\;\; .
\end{equation}
\noindent For this model, the adiabatic speed of sound is,
\begin{equation}
\label{Speed_Sound_Eqn}
c_{s} = \sqrt{\dfrac{K_{s}}{\rho}}
\end{equation}
\noindent where,
%
\begin{equation}
\label{Bulk_Mod_Eqn2}
K_{s} = n \left(\dfrac{\partial P}{\partial n}\right)_{s}
= \dfrac{5}{3}\dfrac{n k_{B}T}{\left(1-b^{\prime} n\right)^{2}} - 2a^{\prime}n^{2}
\end{equation}
\noindent is the bulk modulus. The speed of sound at the critical point is $c_{s,c} = \sqrt{3 k_B T_c/2 m}$. Note that the isothermal (not adiabatic) compressibility diverges at the critical point.

\subsection{Hydrodynamic Equations}

In order to model fluids at the mesoscopic scales, thermal fluctuations have to be incorporated into the continuum Navier-Stokes equations. In fluctuating hydrodynamics this is accomplished by including stochastic fluxes in the equations \cite{Landau_1959}. The full set of compressible fluctuating hydrodynamic equations can be written as,~\cite{EspanolVdW}
\begin{equation}
\label{Continuity_Eqn}
\partial_{t} \rho + \nabla \cdot \left(\rho \mathbf{v}\right) = 0
\end{equation}
\begin{equation}
 \label{Momentum_FNS_Eqn}
\partial_{t} \left(\rho \mathbf{v}\right)
+ \nabla \cdot \left(\rho \mathbf{v} \mathbf{v}^\mathrm{T}\right)
+ \nabla \cdot \ReversibleStress
= \nabla \cdot \left(\StressTensor + \tilde{\StressTensor}\right)
\end{equation}
\begin{equation}
\label{Energy_FNS_Eqn}
\partial_{t} \left(\rho E \right)
+ \nabla \cdot \left(\rho E \mathbf{v} \right)
+ \nabla \cdot \left(\ReversibleStress \cdot \mathbf{v}\right)
= \nabla \cdot \left(\HeatFlux + \tilde{\HeatFlux} \right)
+ \nabla \cdot \left(\left(\StressTensor + \tilde{\StressTensor}\right) \cdot \mathbf{v}\right)
\end{equation}
\noindent where the superscript $\mathrm{T}$ indicates transpose.
The momentum density is $\textbf{g} = \rho \textbf{v}$ and
$\rho E = \frac{1}{2} \rho \textbf{v}^{2} + \rho e$ is the total local energy density.
Note that since $\rho E$ is the local energy density rather than the total energy $U$ (from (\ref{TotalEnergy_Eqn})), there is no interstitial working contribution in the energy flux ~\cite{DunnSerrin,Anderson_1998}.

\indent The deterministic stress tensor has a reversible part, $\ReversibleStress$,
and an irreversible part, $\StressTensor$;
the former can be written as,

\medskip
\begin{equation}
\label{Reversible_Stress_Eqn}
\ReversibleStress = P \IdentityTensor - \KortewegStress
\end{equation}
\medskip

\noindent where the first term is the pressure, $\IdentityTensor$ is the identity tensor and,

\medskip
\begin{equation}
\label{Korteweg_Stress_Eqn}
\KortewegStress = \left[\left(\kappa \rho \nabla^{2} \rho
+ \dfrac{1}{2} \kappa \left|\nabla \rho \right|^{2} \right)
 \IdentityTensor \right]
- \left(\kappa \nabla \rho \otimes \nabla \rho\right)
+ \KortewegCross
\end{equation}
\medskip

\noindent is the Korteweg stress that arises from the gradient energy contribution to the energy.
The first three terms of the Korteweg stress can be derived using variational methods \cite{YangFlemingGibbs,Shang_2011}, or from the GENERIC formulation \cite{EspanolVdW} by applying the standard reversible generator for a single-component fluid (c.f. Eq. (2.60) in \cite{Ottinger_2005}) to \eqref{TotalEnergy_Eqn}. These three terms apply in both isothermal and non-isothermal systems and are widely agreed upon in the literature. The additional cross-coupling term $\KortewegCross$ arises in the presence of temperature gradients, and there is no widely agreed upon form of these terms. In this work we employ the form suggested by Onuki \cite{Onuki_2007},

\medskip
\begin{equation}
\label{Cross_Stress_Eqn}
\KortewegCross = T \rho \nabla \rho \cdot \nabla \left(\dfrac{\kappa}{T}\right) \IdentityTensor
\end{equation}
\medskip

\noindent A cross-coupling term is also present in the work by Dunn \& Serrin \cite{DunnSerrin}; however, the term they introduce has a
different form $\KortewegCross = \rho (\partial\kappa/\partial T) (\nabla \rho \cdot \nabla T)$. If $\kappa$ is assumed to be a constant, as we do in this work, the cross-coupling term proposed by Dunn \& Serrin would vanish. But using Onuki's formulation, the term is still present for constant $\kappa$. Here we chose Onuki's formulation for the cross-coupling term since it is believed to be important in non-equilibrium situations with high temperature gradients \cite{Onuki_2007}.

\indent The deterministic irreversible stress tensor is taken to be that of a Newtonian fluid,

\medskip
\begin{equation}
\label{Stress_Tensor_Eqn}
\StressTensor = \ShearViscosity \left(\nabla \mathbf{v}+(\nabla \mathbf{v})^\mathrm{T}\right)
+ \left( \BulkViscosity - \frac23 \ShearViscosity\right)\left(\nabla \cdot \mathbf{v}\right) \IdentityTensor
\end{equation}
\medskip

\noindent where $\ShearViscosity$ and $\BulkViscosity$ are the shear and bulk viscosities of the fluid, respectively. The deterministic heat flux is given by the Fourier law as
$\HeatFlux = \ThermalConductivity \nabla T$ where $\ThermalConductivity$ is the thermal conductivity.

As discussed in \cite{Landau_1959,Espanol_1998}, the stochastic stress tensor and heat flux can be written as,

\medskip
\begin{equation}
\label{Stochastic_Stress_Eqn}
\tilde{\StressTensor} = \sqrt{2 \ShearViscosity k_{B}T} \, \tilde{\mathcal{W}}^{\textrm{v}} + \left(\sqrt{\dfrac{\BulkViscosity k_{B}T}{3}} - \sqrt{\dfrac{2 \ShearViscosity k_{B}T}{3}} \right) \,
\textrm{Tr}\left(\tilde{\mathcal{W}}^{\textrm{v}}\right)\IdentityTensor
\end{equation}

\medskip
\begin{equation}
\label{Stochastic_Heat_Flux_Eqn}
\tilde{\HeatFlux} = \sqrt{2 \ThermalConductivity k_{B}T^{2}} \, \mathcal{W}^{E}
\end{equation}
\medskip

\noindent where $\tilde{\mathcal{W}}^{\textrm{v}} = (\mathcal{W}^{\textrm{v}}+(\mathcal{W}^{\textrm{v}})^\mathrm{T})/\sqrt{2}$ is a symmetric Gaussian random tensor field and $\mathcal{W}^{\textrm{E}}$ is a Gaussian random vector field. Here $\mathcal{W}^{\textrm{v}}$ and $\mathcal{W}^{\textrm{E}}$ are mutually uncorrelated white
noise Gaussian fields so that

\medskip
\begin{equation}
\langle \mathcal{W}_{ij}^{\textrm{v}} (r,t) \; \mathcal{W}_{kl}^{\textrm{v}}(r',t') \rangle = \delta_{ik}\delta_{jl}\delta(r-r')\delta(t-t')
\end{equation}
\medskip

\begin{equation}
\langle \mathcal{W}_{i}^{\textrm{E}} (r,t) \; \mathcal{W}_{j}^{\textrm{E}}(r',t') \rangle = \delta_{ij}\delta(r-r')\delta(t-t')
\end{equation}
\medskip

\noindent and

\medskip
\begin{equation}
\langle \mathcal{W}_{ij}^{\textrm{v}} (r,t) \; \mathcal{W}_{k}^{\textrm{E}}(r',t') \rangle = 0 \;\;.
\end{equation}
\medskip

\noindent It is important to note that as long as the square gradient term is added to the energetic part of the free energy \cite{EspanolVdW}, as we do in this work, the Korteweg stresses are generated from the reversible part of the dynamics. In this case only the irreversible (dissipative) parts of the momentum and energy flux contribute to the entropy production and thus, by fluctuation-dissipation, there are no surface-tension contributions to the stochastic fluxes.

\section{Spatio-Temporal Discretization}

The numerical integration of the governing equations is based on an extension of the
approach developed in \cite{Donev_2010_2,Balakrishnan2014}.
A finite volume approach is used to discretize the equations spatially. The resulting
system of stochastic ordinary differential equation is then integrated using
a low-storage, total variation diminishing, three-stage Runge-Kutta (RK3)
integrator.
The principal issue in the extension of the methodology in \cite{Balakrishnan2014} is the treatment of
the Korteweg stress term.  Here, we briefly review the discretization used in \cite{Balakrishnan2014}, then
discuss the discretization of $\KortewegStress$ in more detail.
We first write the FNS equations in a compact system form
\begin{equation}
\label{eq:FNS}
\frac{\partial \U}{\partial t} = -
\nabla \cdot \FH - \nabla \cdot \FK - \nabla \cdot \FD - \nabla \cdot \FS  \equiv \mathbf{R}(\U, \mathcal{W})
\end{equation}
\noindent where $\U$ represents the set of conserved hydrodynamic fields:
\begin{equation}
\label{3.2}
\U = \left(\begin{array}{c} \rho \\ \rho \mathbf{v} \\ \rho E \end{array} \right)
\end{equation}
\noindent and  $\FH$, $\FK$, $\FD$, and $\FS$ are the
hyperbolic, Korteweg, diffusive and stochastic flux terms, respectively and $\mathcal{W}(t)$ denotes a collection of independent white-noise processes.
These fluxes may be written as,
\begin{equation}
\FH =
\begin{bmatrix}
\rho \mathbf{v} \\
\rho \mathbf{vv}^T + P\mathbf{I} \\
\mathbf{v} (\rho E + P)
\end{bmatrix}
~ ; \;\;\;
\FK = -
\begin{bmatrix}
0 \\
\KortewegStress \\
\KortewegStress \cdot \mathbf{v}
\end{bmatrix}
~ ; \;\;\;
\FD = -
\begin{bmatrix}
0 \\
\StressTensor \\
{\HeatFlux} + \StressTensor \cdot \mathbf{v}
\end{bmatrix}
~ ; \;\;\;
\FS = -
\begin{bmatrix}
0 \\
\widetilde{\StressTensor}  \\
\widetilde{\HeatFlux} + { \widetilde{\StressTensor}} \cdot \mathbf{v}
\end{bmatrix}.
\end{equation}

\subsection{Spatial Discretization}

The spatial discretization uses finite volume representation with cell spacings in the $x$, $y$ and $z$-directions given by $\Delta x$, $\Delta y$ and $\Delta z$.
$\U_{ijk}$ denotes the average value of $\U$ on cell-$ijk$.
The discretizations are based on centered approximations, designed to ensure that the algorithm satisfies discrete fluctuation-dissipation balance.
See \cite{Balakrishnan2014} for details.

To compute the hyperbolic fluxes, we first compute the primitive variables $Q = \left(\rho,\textbf{v},P,T\right)$ at cell centers.
The primitive variables are then interpolated to faces.
This interpolation can be done using either a simple averaging or
by using a cubic interpolation of $Q$ from cell centers to the faces using the fourth order interpolation formula developed for PPM \cite{Colella_1984},

\begin{equation*}
{Q}_{i + 1/2,j,k} = \dfrac{7}{12}\left({Q}_{i,j,k} + {Q}_{i+1,j,k}\right) - \dfrac{1}{12}\left({Q}_{i-1,j,k} + {Q}_{i+2,j,k}\right)
\end{equation*}
\noindent with similar formulae in the $y$ and $z$ directions.
Numerical tests indicate the the choice of interpolation does not make any significant difference to the results. The hyperbolic fluxes $\FH$ at the faces are then calculated from these interpolants. A key issue here is that the interpolation is based on primitive variables instead of conserved variables.  This is important for multiphase flow because pressure and temperature are smoother at gas / liquid interfaces than the energy density. Using primitive variables thus eliminates the introduction of artifacts from the interpolation process.
These face fluxes are differenced as,
\begin{equation*}
\nabla \cdot \FH \approx \Dftoc \; \FH
\end{equation*}
\noindent where $\Dftoc$ is a discrete divergence operator that computes the cell-centered divergence of a tensor field from values defined at cell faces.

\noindent The heat flux is defined similarly as
\[
\nabla \cdot \HeatFlux = \nabla \cdot (\ThermalConductivity \nabla T ) \approx \Dftoc (\ThermalConductivity \; \Gctof T )
\]
\noindent where $\Gctof$ (which is the adjoint of $\Dftoc$) computes normal components of the gradient at cell faces from cell-centered data and $\ThermalConductivity$ at cell faces is computed by averaging values evaluated at cell centers.

\indent The discretization of the viscous stress terms presents a complication. Standard discretizations of the stress tensor do not satisfy fluctuation dissipation balance. Here we follow the approach described in \cite{Balakrishnan2014} and rewrite the stress tensor as
\begin{equation}
\nabla \cdot \StressTensor =  \nabla \cdot \left( \ShearViscosity \nabla \mathbf{v} \right)+
\nabla \cdot \left [ (\BulkViscosity + \frac{1}{3} \ShearViscosity ) \; \IdentityTensor \; (\nabla \cdot \mathbf{v}) \right]
- \left[ \nabla (\ShearViscosity \;   (\nabla \cdot \mathbf{v}))
- \nabla \cdot (\ShearViscosity (\nabla \mathbf{v})^T) \right].
\label{eq:altstress}
\end{equation}
\noindent We note that the last term in (\ref{eq:altstress}) satisfies
\[
\left[ \nabla  (\ShearViscosity \;   (\nabla \cdot \mathbf{v}))
- \nabla \cdot (\ShearViscosity (\nabla \mathbf{v})^T) \right]
= \left[ (\nabla \ShearViscosity )  ( \nabla \cdot \mathbf{v})
- (\nabla \ShearViscosity) \cdot (\nabla \mathbf{v})^T \right]
\]
\noindent showing that this term vanishes when $\ShearViscosity$ is constant. We will use this alternate form for the discretization, using different approximations for the different terms in equation (\ref{eq:altstress}). For the first term, we approximate
\begin{equation}
\nabla \cdot \left( \ShearViscosity \nabla \mathbf{v} \right)
\approx
\Dftoc \left( \ShearViscosity  \; \Gctof \mathbf{v} \right).
\label{eq:ten1}
\end{equation}
Here, we average adjacent cell-centered values of $\ShearViscosity$ to faces. For the remaining terms we use a nodal (corner) based discretization. For example, we approximate
\begin{equation}
\nabla \left[ (\BulkViscosity + \frac{1}{3} \ShearViscosity ) \; (\nabla \cdot \mathbf{v}) \right]
\approx
\Gntoc \left[ (\BulkViscosity + \frac{1}{3} \ShearViscosity ) \; \Dcton \mathbf{v} \right],
\label{eq:ten2}
\end{equation}
\noindent where $\Dcton$ uses values of a vector field at cell centers to compute the divergence at nodes (corners) and $\Gntoc$ (which is the adjoint of $\Dcton$) computes gradients at cell-centers from values at nodes. Again, the discretizations are standard second-order difference approximations. Here, coefficients are computed by averaging cell-centered values to the corresponding node.

\noindent We also use a nodal discretization for the last terms in (\ref{eq:altstress})
\begin{equation}
\left[ \nabla (\ShearViscosity \;   (\nabla \cdot \mathbf{v}))
- \nabla \cdot (\ShearViscosity (\nabla \mathbf{v})^T) \right]
\approx
\Gntoc \left(  \ShearViscosity \; \Dcton \mathbf{v} \right)
+ \Dntoc \left( \ShearViscosity \; (\Gcton \mathbf{v})^T \right),
\label{eq:ten3}
\end{equation}
where $\Dntoc$ computes divergence at cell centers from nodal values and $\Gcton$ computes nodal gradients from cell-centered values. We note that the second-order derivative terms cancel at the discrete level just as they do in the continuum formulation, leaving only first-order differences when the two terms are combined.

\indent The divergence of the Korteweg stresses with constant gradient energy coefficient can be written as:
\begin{equation}
\label{Div_Korteweg_Eqn}
\nabla \cdot \KortewegStress = \nabla \cdot \left[ \left (\kappa \rho \nabla^{2} \rho + \dfrac{1}{2} \kappa \left|\nabla
\rho \right|^{2} - \frac{\rho \kappa}{T} \nabla \rho \cdot \nabla T \right ) \IdentityTensor \right ] - \nabla \cdot \left(\kappa \nabla \rho \otimes \nabla \rho\right)
\end{equation}
\noindent The four terms in this expression are discretized using a combination of face-centered and nodal discretizations using the discrete operators
defined above. The first term $ \nabla \cdot (\kappa \rho \nabla^{2} \rho \IdentityTensor)$ can be discretized as $\Dftoc [\left(\kappa \rho \mathbf{L} \rho \right)\IdentityTensor]$ where $\textbf{\textrm{L}}$ is the standard 7-point Laplacian (though higher-order or more isotropic discretizations can be used as well),
\begin{equation*}
\left({\mathbf{L}} \rho\right)_{i,j,k} = \dfrac {\left( \rho_{i+1,j,k} - 2\rho_{i,j,k} + \rho_{i-1,j,k}\right)}{(\Delta x)^{2}} +
\dfrac{\left( \rho_{i,j+1,k} - 2\rho_{i,j,k} + \rho_{i,j-1,k}\right)}{(\Delta y)^{2}}
+ \dfrac{\left( \rho_{i,j,k+1} - 2\rho_{i,j,k} + \rho_{i,j,k-1}\right)}{(\Delta z)^{2}}
\end{equation*}
\noindent To compute the first term in (\ref{Div_Korteweg_Eqn}) we first average $\rho {\mathbf{L}} \rho$ to faces
using
\begin{equation*}
\left(\rho \textbf{\textrm{L}} \rho \right)_{i+1/2,j,k} = \dfrac{1}{2}\left(\rho_{i,j,k} \left({\mathbf{L}} \rho\right)_{i,j} +
\rho_{i+1,j,k} \left({\mathbf{L}} \rho\right)_{i+1,j,k}\right)
\end{equation*}
\noindent The discrete divergence operator $\Dftoc$ is then used to compute the divergence of the quantity
$\kappa \rho \textbf{\textrm{L}} \rho \; \IdentityTensor$ at cell centers from values at cell faces.
The second term in Eq.(\ref{Div_Korteweg_Eqn}) is discretized using nodal based operators as follows,
\begin{equation*}
\nabla \cdot \left [ \left(\dfrac{1}{2} \kappa \left|\nabla \rho \right|^{2} \right)
\IdentityTensor \right ] \approx
\Dntoc \left[ \left (\dfrac{1}{2} \kappa \left( \Gcton \rho \right) \cdot \left( \Gcton \rho \right) \right ) \IdentityTensor \right ]
\end{equation*}
\noindent The third term in Eq.(\ref{Div_Korteweg_Eqn}) is also discretized using nodal based operators (for constant $\kappa$) as,
\begin{equation*}
-\nabla \cdot \left [ \left(\dfrac{\rho}{T}\kappa \nabla \rho \cdot \nabla T \right)\IdentityTensor \right ] \approx -\Dntoc \left [ \left(\dfrac{\rho}{T}\kappa \Gcton \rho \cdot  \Gcton T \right)\IdentityTensor \right ]
\end{equation*}
\noindent The last term in Eq.(\ref{Div_Korteweg_Eqn}) can similarly be discretized using the same nodal operators as follows,
\begin{equation*}
\nabla \cdot \left(\kappa \nabla \rho \otimes \nabla \rho\right) \approx
\Dntoc \kappa \left(\Gcton \rho \otimes \Gcton \rho \right)
\end{equation*}
\noindent The term $\nabla \cdot (\KortewegStress \cdot \mathbf{v})$ is computed by averaging nodal
values to faces and forming $\KortewegStress \cdot \mathbf{v}$ at faces and using $\Dftoc$ to difference the face values to cell centers. The term $\nabla \cdot (\StressTensor \cdot \mathbf{v})$ is treated in a similar fashion.

\indent For the noise terms in the energy equation, (\ref{Energy_FNS_Eqn}), the contribution is formed by averaging over the adjacent cells as,
\begin{equation*}
\widetilde{\HeatFlux}_{i+\half,j,k} \approx
\sqrt{2 k_{B} (\ThermalConductivity T^{2})_{i+\half,j,k} } ~ \mathfrak{S} W^{\left(E,x\right)}
\end{equation*}
\noindent where
\begin{equation}
(\ThermalConductivity T^{2})_{i+\half,j,k} =
\half \left( (\ThermalConductivity T^{2})_{i,j,k} + (\ThermalConductivity T^{2})_{i+1,j,k} \right ) \;\;  ,
\label{eq:c2f_averaging}
\end{equation}
\noindent $W^{\left(E,x\right)}$ are normally distributed random numbers and
$\mathfrak{S} = 1/\sqrt{\Delta x \Delta y \Delta z \Delta t}$ to discretize the $\delta$ function correlation in space and time.

\indent The discretization of the noise term in the momentum equation needs to match that of the deterministic stress tensor so that the discretization satisfies fluctuation dissipation balance.
For that reason, we generate noise terms for the first two terms in
(\ref{eq:altstress}) separately.  No stochastic terms are added for the last two
terms because they only involve first derivatives of $\mathbf{v}$.
The stochastic stress tensor is expressed as $\widetilde{\StressTensor}
= \widetilde{\StressTensor}^{(f)} + \widetilde{\StressTensor}^{(n)}$.
The term $\widetilde{\StressTensor}^{(f)}$ corresponds
to the $\nabla \cdot \left( \ShearViscosity \nabla \mathbf{v} \right)$ contribution to the viscous flux; at a face we form it as
\[
\widetilde{\StressTensor}_{i+\half,j,k}^{(f)} =
\sqrt{
2 k_B (\ShearViscosity T)_{i+\half,j,k} } \mathfrak{S} W^{(v,x)},
\]

\noindent where $W^{(v,x)}$ are three-component,
independent face-centered standard Gaussian random variables.
Other faces are treated analogously and the resulting stochastic momentum fluxes are differenced
using the discrete divergence $\Dftoc$.

\noindent The stochastic flux corresponding to the contribution
$ \nabla \left[ (\zeta + \frac{1}{3} \ShearViscosity ) \; (\nabla \cdot \mathbf{v}) \right]$
in the dissipative flux is generated at nodes. Namely,
\[
\widetilde{\StressTensor}_{i+\half,j+\half,k+\half}^{(n)} =
\sqrt{
2 k_B \left [(\zeta + \frac{1}{3}
 \ShearViscosity)T \right ]_{i+\half,j+\half,k+\half} }~\mathfrak{S} W^{(v,n)},
\]
\noindent where $W^{(v,n)}$ are three-component,
independent node-centered standard Gaussian random variables.

\noindent Note that the coefficients at the nodes are averages over the  cells adjacent to the
node, analogous to (\ref{eq:c2f_averaging}).
The divergence of these nodal fluxes is computed using the discrete divergence operator $\Dntoc$.
The viscous heating in the energy equation arising from the stochastic stress is computed analogously
to the deterministic contribution described above.

\subsection{Temporal Discretization}

\indent Following \cite{Donev_2010_2,Balakrishnan2014}, the temporal discretization is based on the three-stage,
low storage total variation diminishing Runge-Kutta (RK3) scheme of Gottlieb and Shu \cite{Gottlieb_1998}.
The three stages of the scheme can be summarized as
\begin{align}
\U^{n+1/3} &= \U^{n} + \Delta t \textbf{\textrm{R}} \left(\U^{n},\W_{1}^{n}\right) \nonumber \\
\U^{n+2/3} &= \dfrac{3}{4}\U^{n} + \dfrac{1}{4}\left(\U^{n+1/3} + \Delta t \textbf{\textrm{R}} \left(\U^{n+1/3},\W_{2}^{n}\right)\right)\nonumber \\
\U^{n+1} &= \dfrac{1}{3}\U^{n} + \dfrac{2}{3}\left(\U^{n+2/3} + \Delta t \textbf{\textrm{R}} \left(\U^{n+2/3},\W_{3}^{n}\right)\right)\label{3.10}
\end{align}
\noindent where $\W_i$ are vectors of Gaussian random variables used in each stage of the integration.
To compute these weights,  at each time step we generate
two vectors of independent normally distributed random variables, $\W^A$ and $\W^B$, and
set
\begin{eqnarray*}
\W_1 &=&  \W^A + \beta_1 \W^B; \\
\W_2 &=&  \W^A +  \beta_2  \W^B; \\
\W_3 &=&  \W^A+  \beta_3  \W^B,
\end{eqnarray*}
\noindent where $\beta_1 = (2 \sqrt{2}+ \sqrt{3})/5 $, $\beta_2 = (-4 \sqrt{2}+ 3 \sqrt{3})/5$, and
$\beta_3 = (\sqrt{2} - 2 \sqrt{3})/10$.
With this choice of weights the temporal integration is weakly second-order accurate for additive noise, third-order accurate for equilibrium covariances (static structure factors) \cite{Delong_2013}, and third-order deterministically.

\subsection{Stability}

The numerical scheme described above is fully explicit and is only stable provided $\Delta t$ is sufficiently small.
We can estimate bounds on the time step by substituting Fourier modes into the linearized equations.
The resulting amplification matrix includes real and imaginary components.
The imaginary component corresponds to the reversible component of the
dynamics given by the hyperbolic fluxes and the Korteweg stress.
The viscous stress tensor and the heat flux correspond to the real component.

For stability of the reversible dynamics, we require that
\begin{equation}
\Delta t_r \sum_{\alpha=x,y,z} \frac{|u_\alpha|+ \sqrt{c_s^2 + 2 \kappa \rho \mathcal{V}}}{\Delta \alpha} \leq C
\end{equation}
\indent where
\[
\mathcal{V} = \frac{2}{\Delta x^2}+\frac{2}{\Delta y^2}+ \frac{2}{\Delta z^2}
\]
\noindent is the absolute value of the diagonal element of the standard discrete Laplacian
and $C$ is a constant characterizing stability of the Runge-Kutta scheme ($C \simeq 1.7$).

For the irreversible dynamics given by the dissipative terms we require that
\begin{equation}
\Delta t_i \frac{(\frac{4}{3} \eta + \zeta)}{\rho} \mathcal{V} \leq C \;\;\; \mathrm{and} \;\;\;
\Delta t_i \frac{2 \lambda m}{3 \rho  k_B} \mathcal{V} \leq C
\end{equation}

\noindent To compute the maximum allowable time step we compute the minimum stable values of $\Delta t_r$ and $\Delta t_i$ over
all the cells using the local properties and then multiply by a safety factor less than unity.  We also restrict the time step
to be less than or equal to a prescribed value set at run time.  For the simulations presented below, we typically
take a time step of 10\% of the maximum in order to minimize temporal discretization errors and thus focus our attention on the spatial discretization.

\section{Numerical Validation}

\indent The algorithm presented in the previous section has been extensively validated for the case of a single phase fluid.~\cite{Donev_2010_2,Balakrishnan2014} As such we focus on validation of the code with the addition of Korteweg stresses.
As a prelude to that discussion we first consider the choice of the gradient energy coefficient, $\kappa$. All other physical parameters used in the simulations are given in Table \ref{tab:1}; these parameters are treated as constants unless otherwise stated.

\begin{table}[!htb]
\begin{center}
  \begin{tabular} {| c | c |}
    \hline
    Parameter & Value \\ \hline
    van der Waals parameter, $a^{\prime}$ & 3.736 x 10\textsuperscript{-36} $~\mathrm{erg}/~\mathrm{cm}^{3}$\\
    van der Waals parameter, $b^{\prime}$ & 5.315 x 10\textsuperscript{-23} $~\mathrm{cm}^{3}$\\
    Critical density, $\rho_{c}$ & 0.415995 $~\mathrm{g}/\mathrm{cc}$\\
    Critical temperature, $T_{c}$ & 150.85 $~\mathrm{K}$\\
    Critical temperature, $P_{c}$ & 0.4897835 $~\times 10^{8}~\mathrm{dynes}/\mathrm{cm}^{2}$\\
    Gradient energy coefficient, $\kappa$ & 1.24 x 10\textsuperscript{-5} $~\mathrm{cm}^{7}/(~\mathrm{g}~\mathrm{s}^{2})$\\
    Molecular mass, $m \, N_A$ & 39.948 $~\mathrm{amu}$\\
    Shear viscosity, $\ShearViscosity$ & 5.347 x 10\textsuperscript{-4} $~\mathrm{g}/(~\mathrm{cm}~\mathrm{s})$\\
    Bulk viscosity, $\BulkViscosity$ & 0 $~\mathrm{g}/(~\mathrm{cm}~\mathrm{s})$\\
    Thermal conductivity, $\ThermalConductivity$ & 0.05463 x 10\textsuperscript{5} $~\mathrm{erg}/(~\mathrm{cm}~\mathrm{s}~\mathrm{K})$ \\
    \hline
  \end{tabular}
  \caption{Physical parameters used in the simulations.
The fluid is similar to Argon near its critical point~\cite{Weast_1972};
the values of shear viscosity and thermal conductivity
are for Argon at $T=145$~K, $P = 4.0 \times 10^5~\mathrm{dynes/cm}^2$ \cite{Hanley_1974}.}
  \label{tab:1}
\end{center}
\end{table}

\subsection{Estimating the Gradient Energy Coefficient}

\indent
The theory by Rayleigh relates $\kappa$ to the attractive part of the inter-molecular potential $\phi^{a}(r)$ \cite{Rowlinson_2002}, specifically,
\begin{equation}
\label{Ralyeigh_Kappa_Eqn}
\kappa = -\dfrac{1}{6 m^{2}}\int \, r^{2} \phi^{a}\left(r\right) \textrm{d}\textbf{r}
\end{equation}
From the statistical mechanics of liquids it is known that, in general, the relation between $\kappa$ and the potential is more complicated as it depends on the direct correlation function of the fluid \cite{Rowlinson_2002,Evans_1979,Evans_1992}. The expression in (\ref{Ralyeigh_Kappa_Eqn}) is obtained from the Percus-Yevick approximation in the limit of a weak attractive potential ($\phi^{a} \ll kT$) and a pair correlation function of unity.
To estimate $\kappa$ and its relation to the van der Waals parameters $a'$ and $b'$, we introduce the Sutherland model for the inter-molecular potential,
\begin{equation}
\phi(r) = \left\{
\begin{array}{cc}
  \infty & r < r_0 \\
  \phi^{a}(r) & r \geq r_0
\end{array}
\qquad\mathrm{where}\qquad
\phi^{a}(r) = -4 \epsilon \left(\dfrac{r_0}{r}\right)^{6}
\right.
\end{equation}
In the low-density approximation, $b^{\prime} = (2/3)\pi r_0^3$ and $a^{\prime} = 4\epsilon b^{\prime}$~\cite{Pathria_2007}; from (\ref{Ralyeigh_Kappa_Eqn}) we have,
\begin{equation}
\label{LJ_Kappa_Eqn}
\kappa_{HS} = \dfrac{8\pi\epsilon r_0^{5}}{3 m^{2}}
\end{equation}

For the values of $a^{\prime}$ and $b^{\prime}$ used in our simulations
the corresponding well depth and radius for $\phi^{a}(r)$ are: $\epsilon = 1.76 \times 10^{-28}~\mathrm{erg}$, $r_0 = 2.94 \times 10^{-8}~\mathrm{cm}$, which are very close to the standard  Lennard-Jones parameters for Argon ($\epsilon = 1.65 \times 10^{-28}~\mathrm{erg}$, $r_0 = 3.4 \times 10^{-8}~\mathrm{cm}$). From (\ref{LJ_Kappa_Eqn}) the corresponding value of the gradient energy coefficient is $\kappa_{HS} = 0.736 \times 10^{-6}~\mathrm{cm}^{7}/\mathrm{g}\cdot\mathrm{s}^{2}$, which may be considered as a lower-bound for $\kappa$ given the approximations in the Rayleigh theory.

Although this estimate provides a lower bound, we do not know {\it {a priori}} whether it
is an accurate estimate for $\kappa$.
To obtain an estimate of $\kappa$, we performed
a series of calculations to measure the relationship between $\kappa$ and
the surface tension, $\sigma$, for a flat interface and compared $\sigma$ with experimental observations.
The surface tension can be calculated from the density gradient of a perfectly flat infinite interface as,

\begin{equation}
\label{sigma_estimate}
\sigma = \kappa \int_{y_\mathrm{g}}^{y_\mathrm{\ell}} \left( \frac{d\rho}{dy}\right)^2 dy
\end{equation}

\noindent where $y_\mathrm{g}$ and $y_\mathrm{\ell}$ are points in the pure gas and liquid regions, respectively.
For argon near the critical point, experimental measurements show that $\sigma \approx 0.57 \mathrm{dynes}/\mathrm{cm}^2$ \cite{Kuz_1987} at $T=145$~K where $T_{c} = 150.85 K$.

\begin{figure}
  \setlength{\belowcaptionskip}{5pt}
  \centering
  \includegraphics[width=0.6\textwidth]{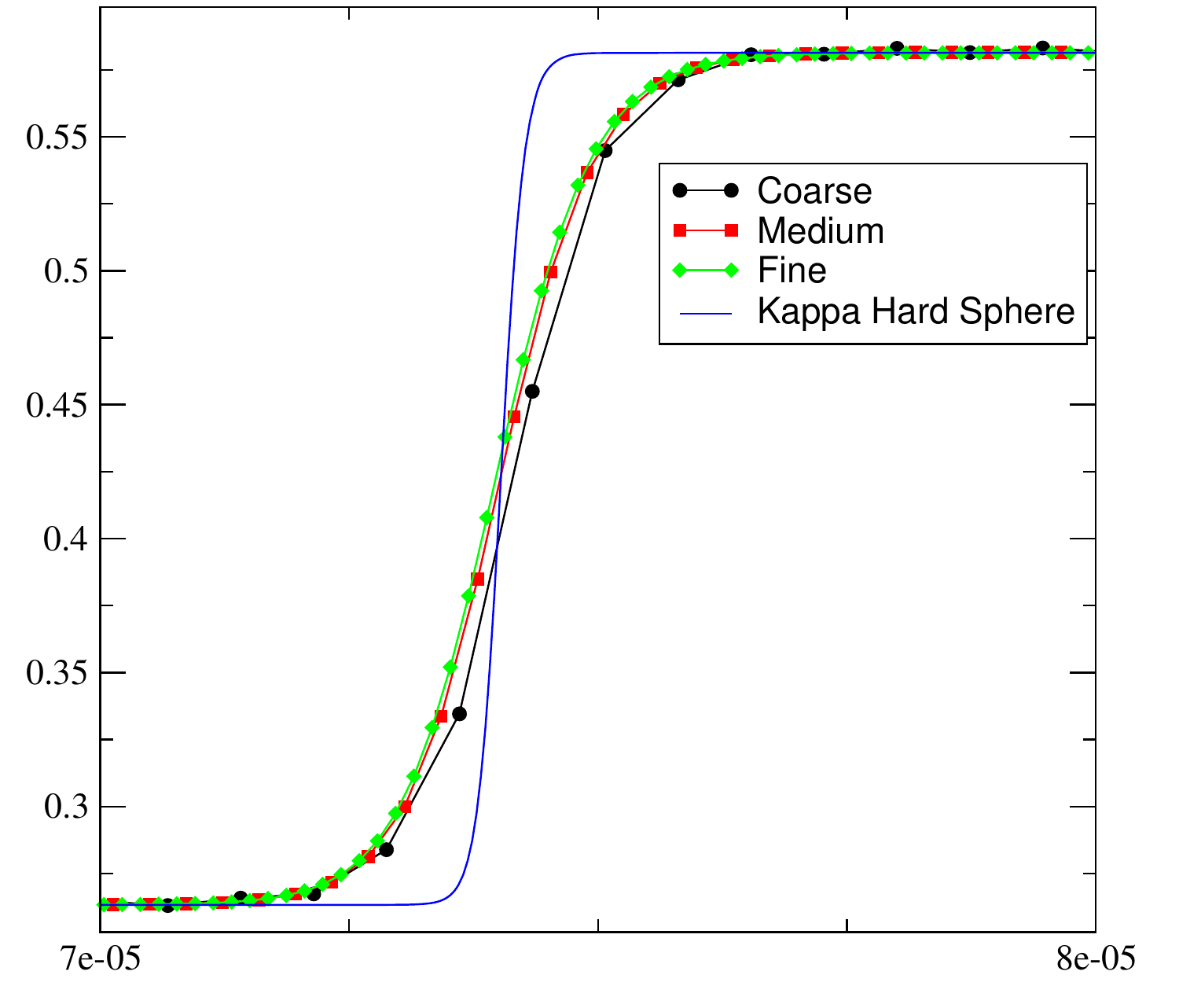}
  \caption{Density variation across the liquid-vapor interface for $\kappa_{HS} = 0.736 \times 10^{-6}~\mathrm{cm}^{7}/\mathrm{g}\cdot\mathrm{s}^{2}$ (solid line) and for $\kappa = 1.24 \times 10^{-5}~\mathrm{cm}^{7}/\mathrm{g}\cdot\mathrm{s}^{2}$ at different resolutions (symbols).} 
  \label{interface}
\end{figure}

Deterministic simulations were performed on a periodic domain in 1D with a slab of liquid in the center. The temperature was set at $T = T_{c} - 5 K$ with the liquid density set at 0.561 g/cc and vapor density at 0.256 g/cc.
Resolution studies were conducted to compute converged estimates of $\sigma$ for given values of $\kappa$ from the steady state profile of the density via \eqref{sigma_estimate}.
Our initial numerical tests revealed that $\kappa_{HS}$ gave extremely sharp interfaces as shown in Fig.\ref{interface}. The calculated value of $\sigma$ was more than a factor of four too small compared with the experimental value. After a number of trials, we settled on a value of $\kappa$ approximately one order of magnitude larger, specifically, $\kappa = 1.24 \times 10^{-5}~\mathrm{cm}^{7}/\mathrm{g}\cdot\mathrm{s}^{2}$.

For a system of $256$ cells with a liquid slab that was initially 60 cells wide, using $\Delta x = 4.6875 \times 10^{-7}~\mathrm{cm}$ the estimated surface tension was
$\sigma = 0.62 \; \mathrm{dynes}/\mathrm{cm}^2$,
comparable to experimental measurements.
Simulations with larger system sizes and larger slabs with the same mesh spacing gave results that differed by about 2\%.
More detailed refinement studies indicate that the interface is slightly under-resolved (see Fig. \ref{interface}); for finer grid spacings the surface tension converges to approximately $\sigma = 0.68 \;  \mathrm{dynes}/\mathrm{cm}^2$. However, at these finer grids the cell volumes are so small that the stochastic fluxes, when included,
have large amplitudes, potentially leading to numerical instabilities. This problem is exacerbated by the presence of a gas phase which has much larger compressibility and thus larger density fluctuations than the liquid phase, especially close to criticality. Here we use a slightly under-resolved mesh spacing for the simulations reported below. Some {\em heuristic} corrections have been proposed in Appendix B of \cite{Shang_2011} to prevent numerical instabilities due to large density fluctuations. A more systematic alternative is to adjust the magnitude of the fluctuations by filtering the stochastic fluxes (see Appendix B in \cite{LowMachExplicit}); this corresponds to separating the grid spacing (and thus the spatial discretization of various differential operators) from the physical coarse-graining length scale.

\subsection{Structure Factor Calculations}

To validate the FNS algorithm in two and three dimensions, static structure factor calculations for density are performed in pure gas and liquid regions and compared with analytical expressions. The parameters used in the simulations are listed in Table \ref{tab:1}. The static structure factor of density is calculated as,
\begin{equation}
\label{Structure_Factor_Eqn}
\textrm{S}\left(\textbf{k}\right) = \langle \delta \hat{\rho} \left(\textbf{k},t\right)
\delta \hat{\rho}^* \left(\textbf{k},t\right) \rangle
\end{equation}
where the Fourier transform of the density fluctuations is
\begin{equation}
\label{Density_Transform_Eqn}
\delta \hat{\rho}\left(\textbf{k},t\right) = \int \textrm{d}\textbf{r} \, e^{-i\textbf{k} \cdot \textbf{r}} \left(\rho\left(\textbf{r},t\right) - \bar{\rho}\right)
\end{equation}
with $\bar{\rho}$ being the mean density and $\langle \cdots \rangle$ denoting a time average over the length of the simulation.

For the analytical calculations, the linearized fluctuating Navier-Stokes equations are rewritten as an Ornstein-Uhlenbeck process and the steady-state covariances (static structure factor) are calculated from the fluctuation-dissipation balance \cite{Donev_2010_2},
\begin{equation}
\label{Structure_FluctDisp_Eqn}
\textrm{S}\left(\textbf{k}\right) = \dfrac{\rho^{2}k_{B}T}{K_{T} + \rho^{2}\kappa \textbf{k}^{2}}
\end{equation}

\noindent where $K_{T} = \rho \left(\dfrac{\partial P}{\partial \rho}\right)_{T}$ is the isothermal bulk modulus given by,
\begin{equation*}
K_{T} = \dfrac{nk_{B}T}{\left(1-\dfrac{n}{3n_{c}}\right)} - \dfrac{9}{4}\dfrac{k_{B}T_{c}n^{2}}{n_{c}} + \dfrac{1}{3}\dfrac{k_{B}Tn^{2}}{n_{c}\left(1-\dfrac{n}{3n_{c}}\right)^{2}}
\end{equation*}
where the critical parameters are given in Eq.(\ref{VdW_Critical_Eqn}).

At $\textbf{k} = 0$, the standard expression for static structure factor of density is recovered \cite{Landau_80} and we see that the effect of the Korteweg stresses is to make the structure factor wavenumber-dependent. In the discrete approximation, the $\textbf{k}^{2}$ term in (\ref{Structure_FluctDisp_Eqn}) comes from the Laplacian term in (\ref{Korteweg_Stress_Eqn}). We used the standard 7 point Laplacian to discretize this term. For the purpose of comparison with the simulation results, we replace the wavenumber squared $\textbf{k}^{2}$ in the denominator of Eq.(\ref{Structure_FluctDisp_Eqn}) with a modified wavenumber that comes from the Fourier transform of the discrete Laplacian operator,

\begin{equation*}
\textbf{k}^{2} \approx
\dfrac{\textrm{exp}(ik_{x}\Delta x)-2+\textrm{exp}(-ik_{x}\Delta x)}{\left(\Delta x/2\right)^{2}} + \dfrac{\textrm{exp}(ik_{y}\Delta y)-2+\textrm{exp}(-ik_{y}\Delta y)}{\left(\Delta y/2\right)^{2}} + \dfrac{\textrm{exp}(ik_{z}\Delta z)-2+\textrm{exp}(-ik_{z}\Delta z)}{\left(\Delta z/2\right)^{2}} \;\;
\end{equation*}

\noindent which simplifies to,

\begin{equation}
\label{k_discrete}
\textbf{k}^{2} \approx
\dfrac{\textrm{sin}^{2}\left(k_{x} \Delta x/2\right)}{\left(\Delta x/2\right)^{2}} + \dfrac{\textrm{sin}^{2}\left(k_{y} \Delta y/2\right)}{\left(\Delta y/2\right)^{2}} + \dfrac{\textrm{sin}^{2}\left(k_{z} \Delta z/2\right)}{\left(\Delta z/2\right)^{2}} \;\;
\end{equation}

\begin{figure}
  \setlength{\belowcaptionskip}{5pt}
  \centering
  \begin{subfigure}[b]{1.0\textwidth}
    \centering
    \includegraphics[width=0.5\textwidth]{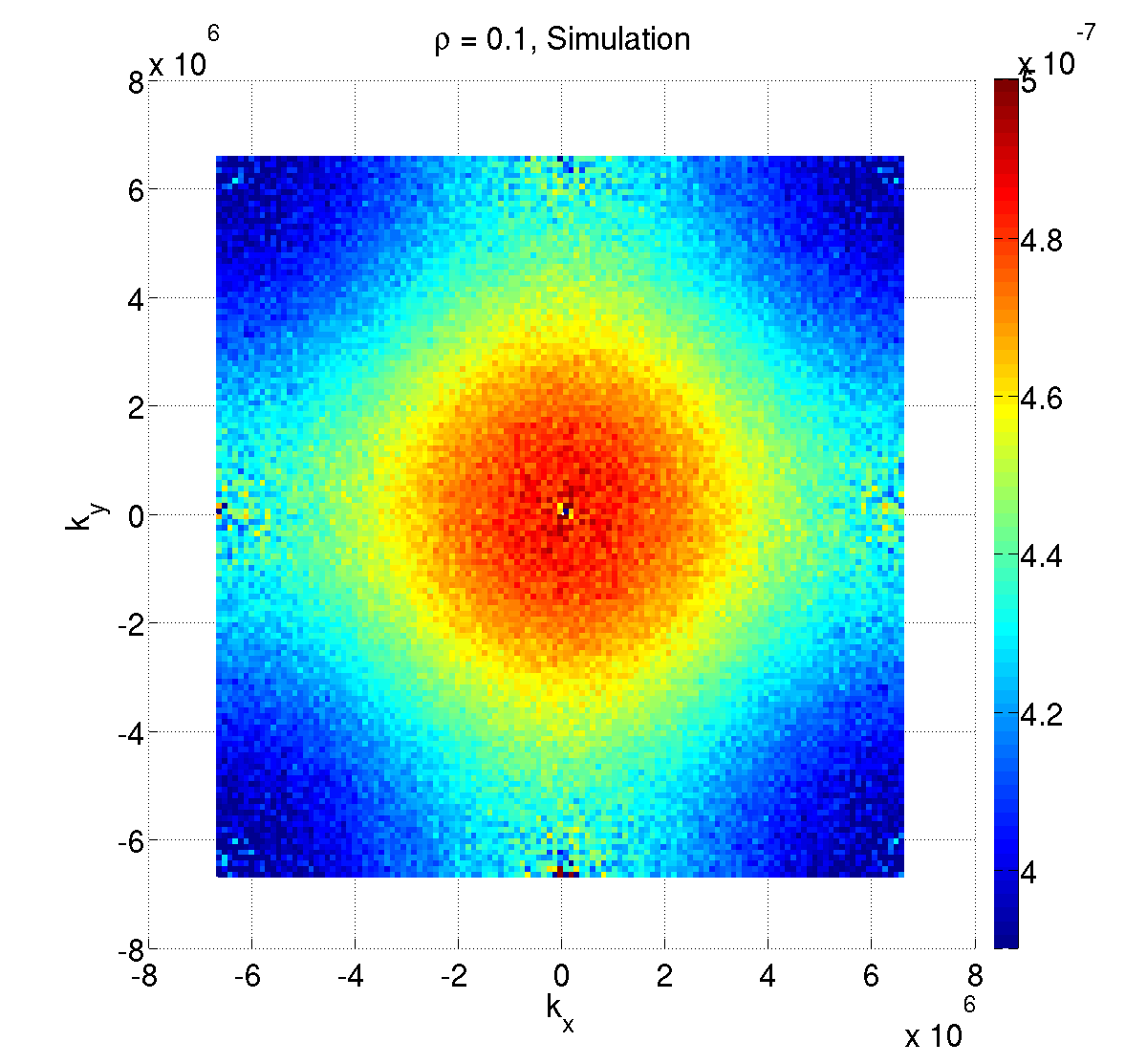}
    \caption{}
    \label{2DSFac_Gas_Sim}
  \end{subfigure}
  \begin{subfigure}[b]{1.0\textwidth}
    \centering
    \includegraphics[width=0.5\textwidth]{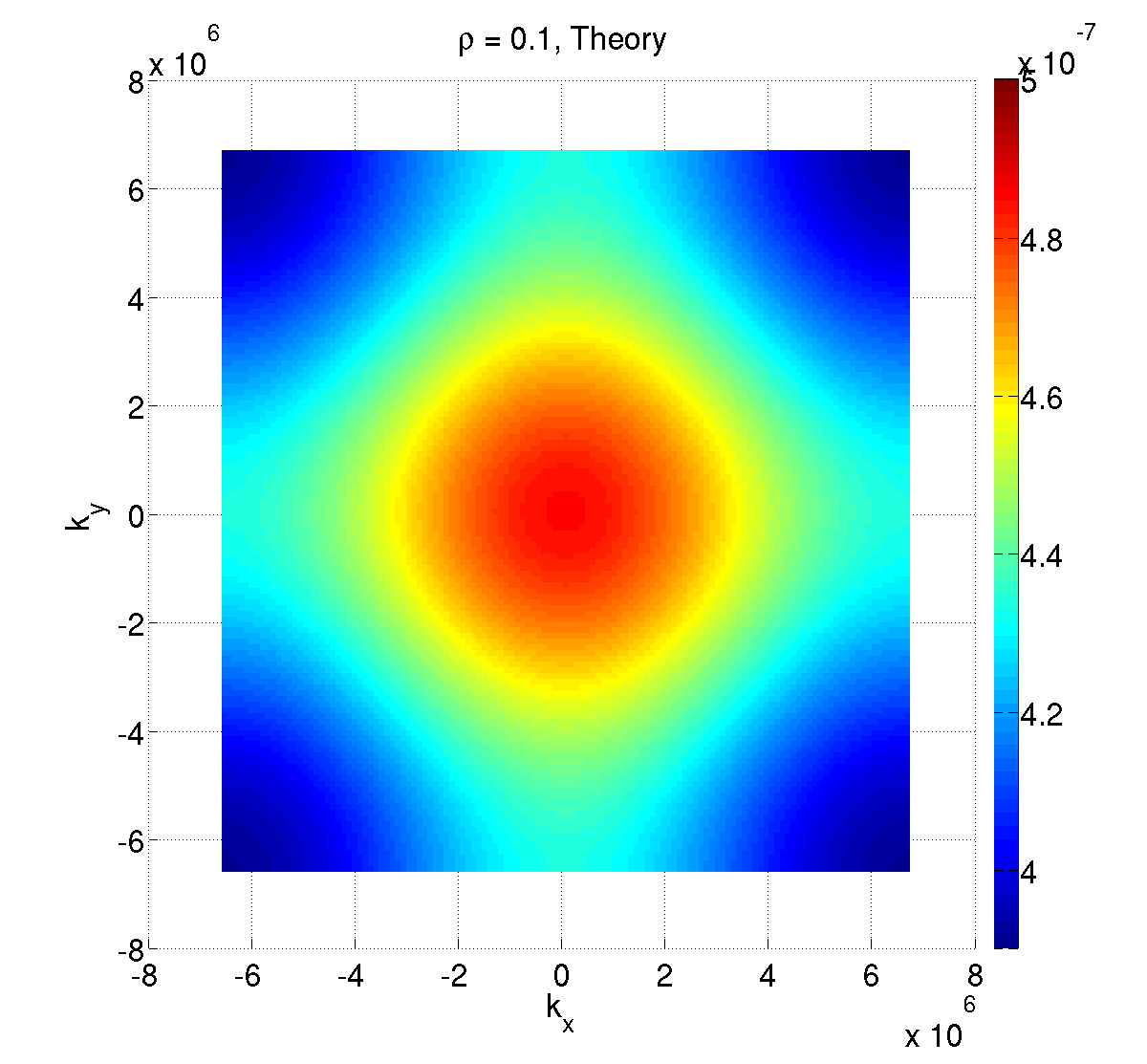}
    \caption{}
    \label{2DSFac_Gas_Theory}
  \end{subfigure}
  \caption{2D structure factor $S\left(\textbf{k}\right)$ for density in the pure gas region at $\rho = 0.1 ~\mathrm{g/cc}$, $T = 145.85~\mathrm{K}$, (a) simulation results, (b) analytical calculations;  system volume = $36 \times 10^{-14}~ \mathrm{cm}^{3}$; periodic boundary conditions with 128\textsuperscript{2} grid points; simulation results have been averaged over $2 \times 10^{7}$ time steps.}\label{2DSFac_Gas}
\end{figure}

The system is prepared initially in a sub-critical state at temperature $T = T_{c} - 5 ~\mathrm{K}$ and at a density either in the gas phase ($\rho = 0.1~\mathrm{g/cc}$)
or in the liquid phase ($\rho = 0.63~\mathrm{g/cc}$). The system size used in 2D simulations is $\left(6\textrm{x}10^{-5}\right)^{2}~\mathrm{cm}^{2}$ with a thickness of $1 \times 10^{-4}~\mathrm{cm}$ using a $128 \times 128$ grid. The time step used is $1.0 \times 10^{-13}~\mathrm{s}$ with a single random variable each time step. In Figs. \ref{2DSFac_Gas_Sim} and \ref{2DSFac_Gas_Theory}, the 2D simulation results for the static structure factor of density is compared with theory for the case where the initial state in the pure gas regime. Good agreement is found for most values of wavenumber. For low $\textbf{k}$ (longest, slowest modes), the structure factor takes a very long time to converge and hence the values at the center of the k-grid in Fig. \ref{2DSFac_Gas_Sim} show the largest errors.


\begin{figure}
  \setlength{\belowcaptionskip}{5pt}
  \centering
  \begin{subfigure}[b]{1.0\textwidth}
    \centering
    \includegraphics[width=0.5\textwidth]{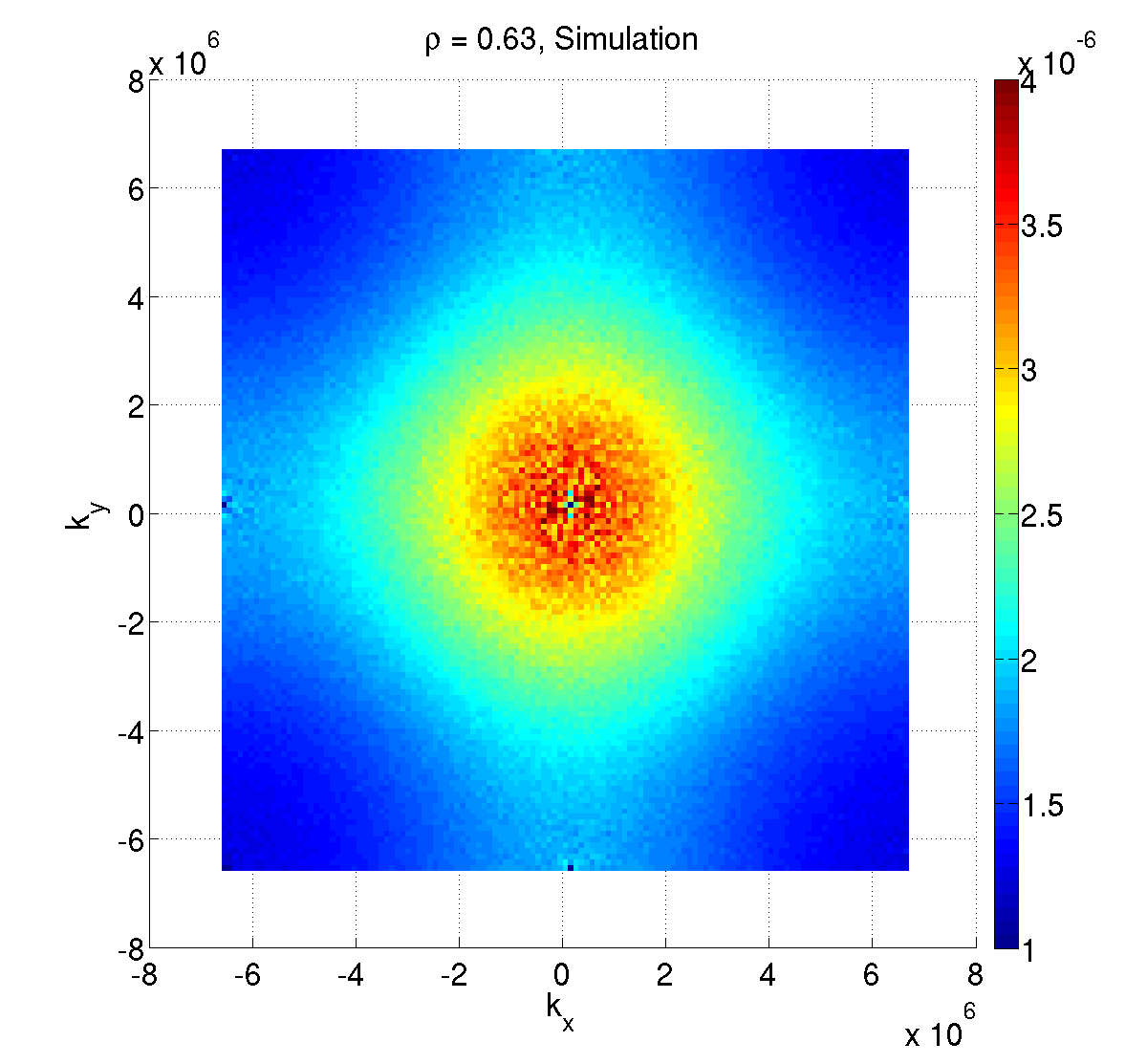}
    \caption{}
    \label{2DSFac_Liquid_Sim}
  \end{subfigure}
  \begin{subfigure}[b]{1.0\textwidth}
    \centering
    \includegraphics[width=0.5\textwidth]{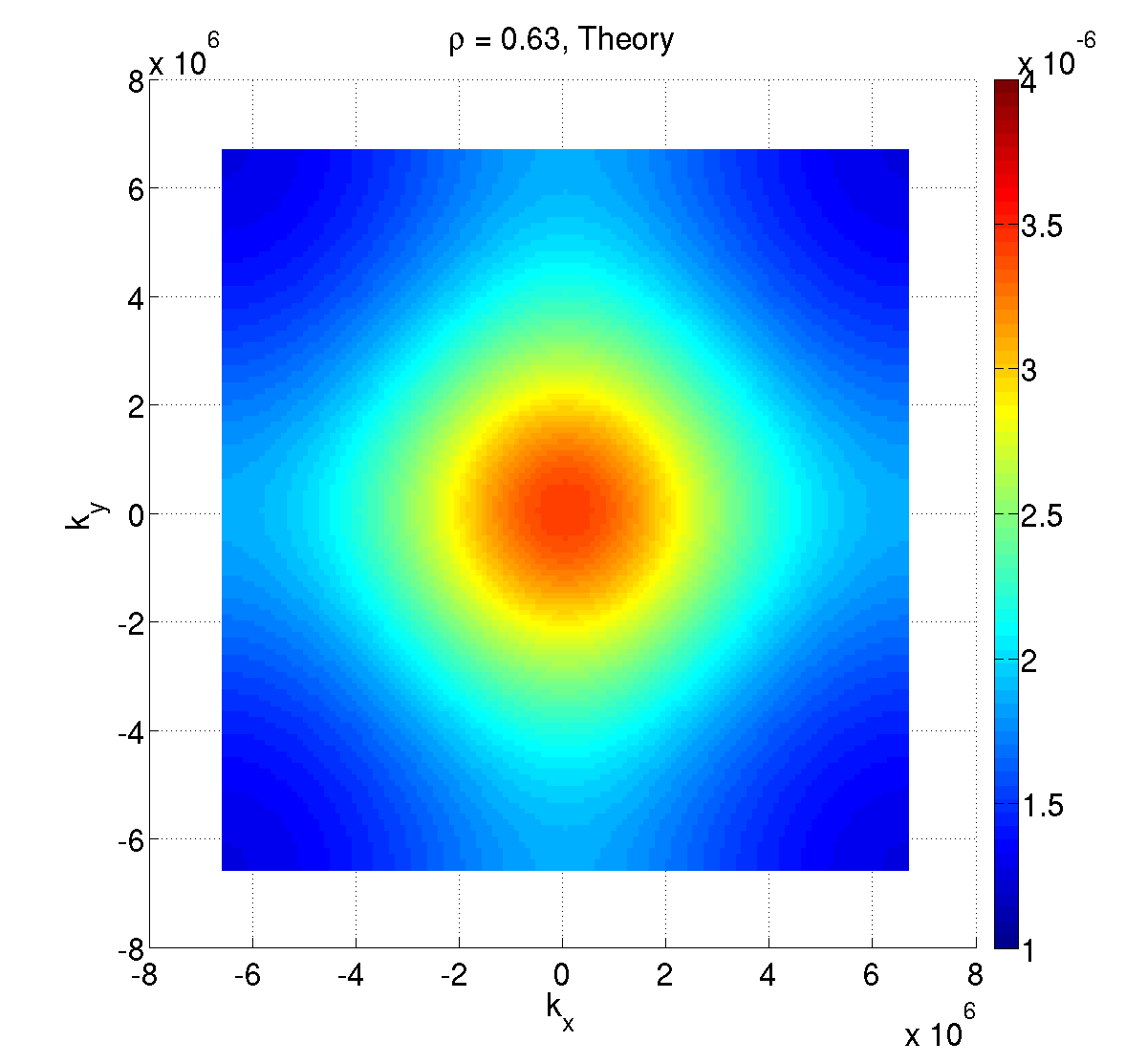}
    \caption{}
    \label{2DSFac_Liquid_Theory}
  \end{subfigure}
  \caption{2D structure factor $S\left(\textbf{k}\right)$ for density in the pure liquid region at $\rho = 0.63 ~\mathrm{g/cc}$, $T = 145.85 ~\mathrm{K}$, (a) simulation results, (b) analytical calculations; system volume = $36 \times 10^{-14}~ \mathrm{cm}^{3}$; periodic boundary conditions with 128\textsuperscript{2} grid points; simulation results have been averaged over $4.9 \times 10^{6}$ time steps.}\label{2DSFac_Liquid}
\end{figure}

\noindent In the case when the initial state is in the pure liquid regime, the simulation results in Fig. \ref{2DSFac_Liquid_Sim} are again in very good agreement with the theoretical prediction shown in Fig. \ref{2DSFac_Liquid_Theory}. We note in numerical structure factors in liquid are smoother
than in the gas since density fluctuations
are smaller in relative magnitude due to the lower compressibility of the liquid state.

\begin{figure}
  \setlength{\belowcaptionskip}{5pt}
  \centering
  \begin{subfigure}[b]{1.0\textwidth}
    \centering
    \includegraphics[width=0.5\textwidth]{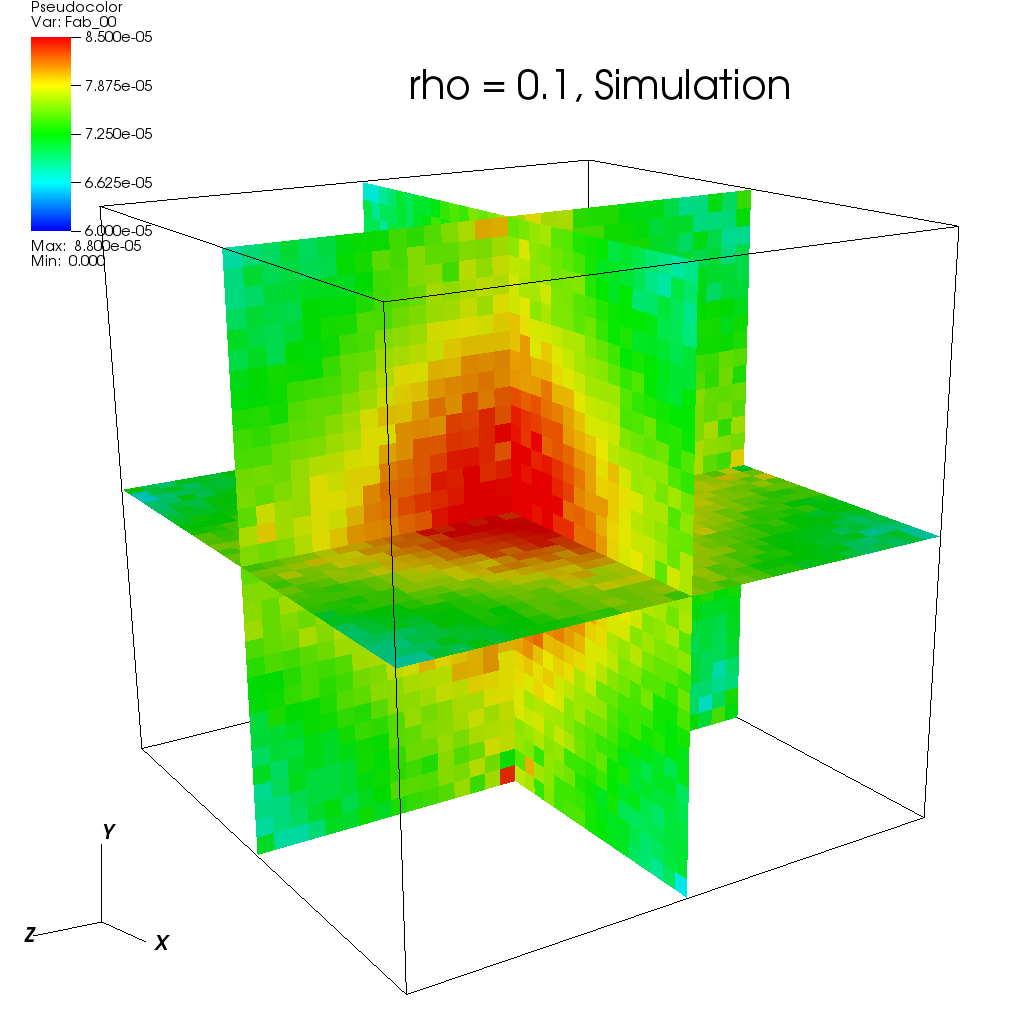}
    \caption{}
    \label{3DSFac_Gas_Sim}
  \end{subfigure}
  \begin{subfigure}[b]{1.0\textwidth}
    \centering
    \includegraphics[width=0.5\textwidth]{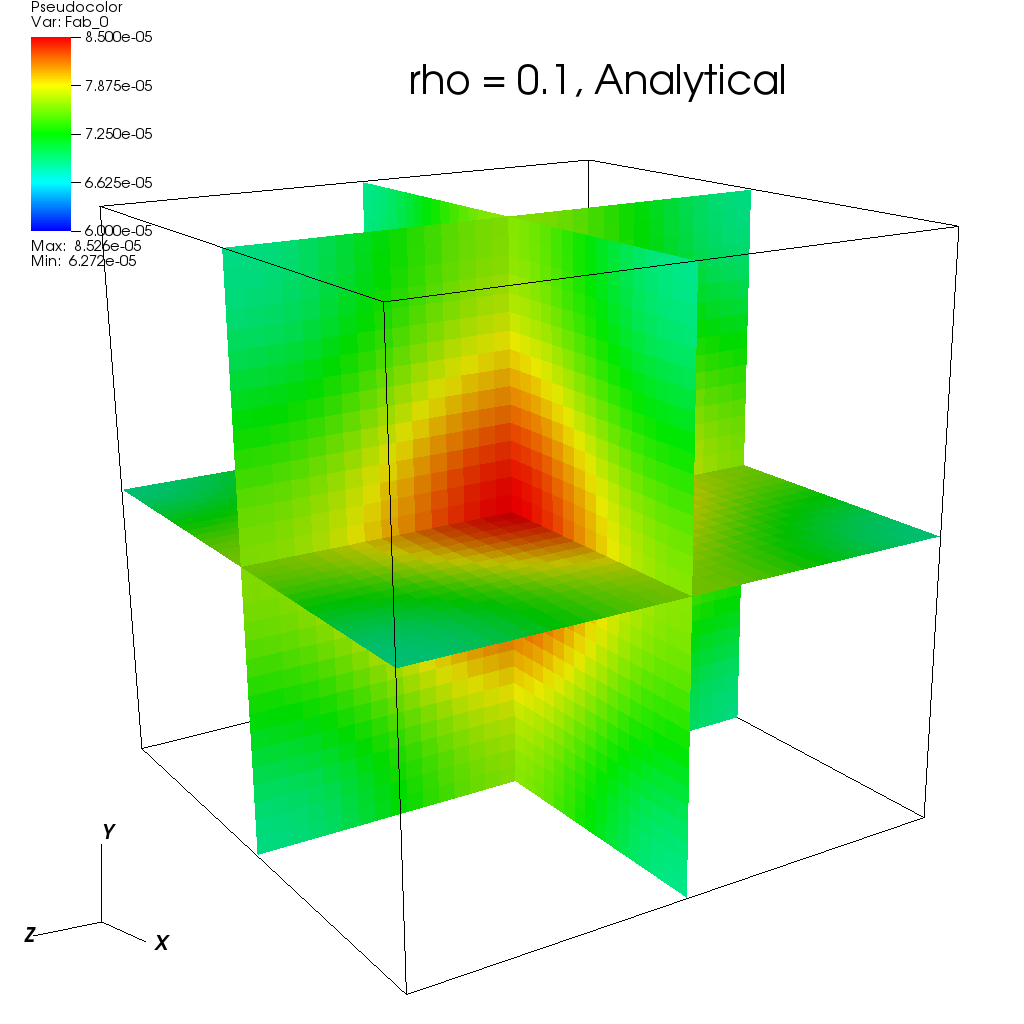}
    \caption{}
    \label{3DSFac_Gas_Theory}
  \end{subfigure}
  \caption{3D structure factor $S\left(\textbf{k}\right)$ for density in the pure gas region at $\rho = 0.1 ~\mathrm{g/cc}$, $T = 145.85 ~\mathrm{K}$, (a) simulation results, (b) analytical calculations; periodic boundary conditions with $32^{3}$ grid points; simulation results have been averaged over
$4 \times 10^{6}$ time steps.}\label{3DSFac_Gas}
\end{figure}

\begin{figure}
  \setlength{\belowcaptionskip}{5pt}
  \centering
  \begin{subfigure}[b]{1.0\textwidth}
    \centering
    \includegraphics[width=0.5\textwidth]{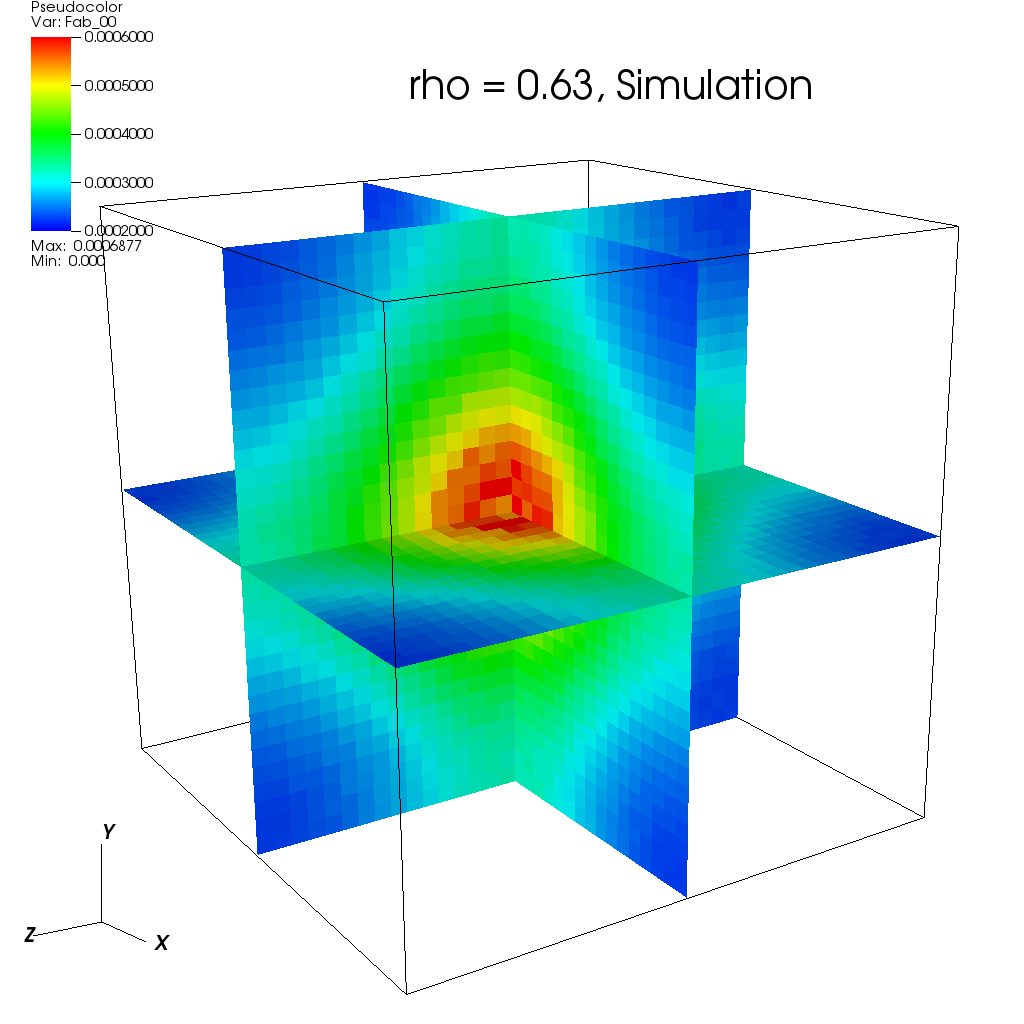}
    \caption{}
    \label{3DSFac_Liquid_Sim}
  \end{subfigure}
  \begin{subfigure}[b]{1.0\textwidth}
    \centering
    \includegraphics[width=0.5\textwidth]{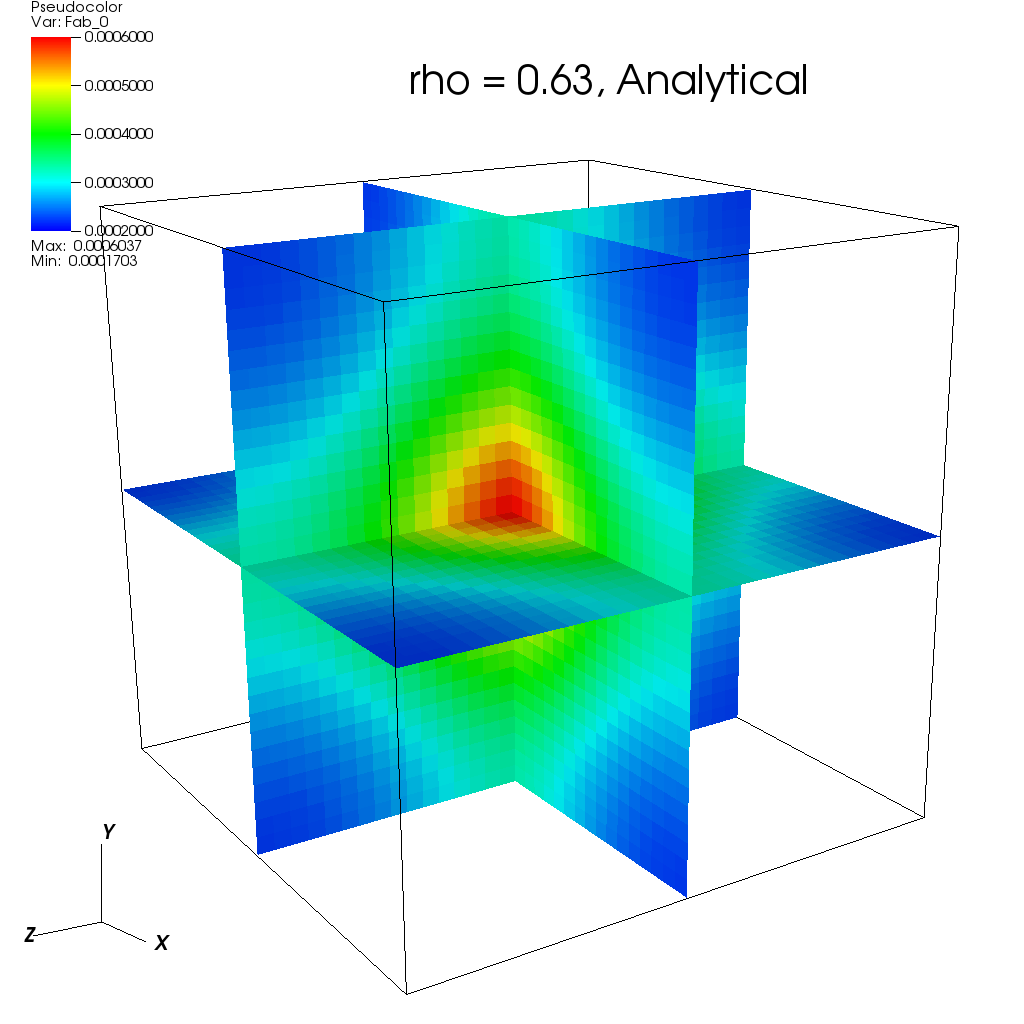}
    \caption{}
    \label{3DSFac_Liquid_Theory}
  \end{subfigure}
  \caption{3D structure factor $S\left(\textbf{k}\right)$ for density in the pure liquid region at $\rho = 0.63 ~\mathrm{g/cc}$, $T = 145.85 ~\mathrm{K}$, (a) simulation results, (b) analytical calculations; periodic boundary conditions with 32\textsuperscript{3} grid points; simulation results have been averaged over $4 \times 10^{6}$ time steps.}\label{3DSFac_Liquid}
\end{figure}

Analogous simulations were run in 3D on a triply periodic domain
with dimensions $L_x = L_y = L_z = 1.6 \times 10^{-5}~\mathrm{cm}$ discretized
using a $32 \times 32 \times 32$ grid.
The time step was $\Delta t = 5.0 \times 10^{-13}~\mathrm{secs}$, corresponding to  approximately
10\% of the maximum allowable time step.
For these simulations, an initial $4\times 10^4$ steps were taken to allow the fluctuations to reach equilibrium.  Data was then taken
every 10 steps for $4\times 10^6$ steps to compute the static structure factors. The figures for the 3D structure factors, Fig. \ref{3DSFac_Gas} \& Fig. \ref{3DSFac_Liquid}, reveal a similar pattern as the 2D structure factors. In Fig. \ref{3DSFac_Gas_Sim}, the deviations from theoretical calculations in Fig. \ref{3DSFac_Gas_Theory}
are larger at the center of the k-grid as a result of slow convergence of long wavelength modes.

\indent In the pure liquid region, the simulations results in Fig. \ref{3DSFac_Liquid_Sim} again
compare very well with theoretical results in Fig. \ref{3DSFac_Liquid_Theory}. Overall, the 2D and 3D structure factors calculated from simulations are in very good agreement with analytical calculations, which serves as a good validation of the numerical scheme.
(We note that additional tests were run in 3D using transport properties proportional to density, i.e.,
$\ShearViscosity = \ShearViscosity_c \rho / \rho_c$, etc. and with $\zeta = \ShearViscosity$.
The results, which are not presented here, are essentially unchanged from the constant coefficient results.)

\subsection{Liquid-Vapor Fluctuating Interfaces}

A fluid interface near a critical point is often imagined to be an interface distorted due to thermal fluctuations.
Distortions of the normal interface are often characterized as capillary waves.
Capillary wave theory \cite{Buff_1965,Rowlinson_2002,Abraham_1979} predicts that, at thermodynamic equilibrium, the variance of the interface height varies with the wavenumber as $k^{-2}$ in the absence of gravitational attraction.
For a smooth, nearly flat interface, a height function, $h(x)$, can be defined that measures the deviations of the interface from the flat profile.
According to capillary wave theory, the variance of the height fluctuations in the Fourier domain can be written as,

\medskip
\begin{equation}
\label{SurfaceTension_CWT}
\langle \left | \hat{h}_{k} \right |^{2} \rangle = \dfrac{k_{B}T}{Ak^{2}\sigma}
\end{equation}
\medskip

\noindent where $A$ is the area of the interface, $k$ is the wavenumber and $\langle \cdots \rangle$ denotes the time average.
The height function is defined as follows: as the liquid layer is approached from the vapor layer,
find the first $y$ location
where the linear interpolant of the cell-centered values
gives $ \rho = \frac12 (\rho_{\ell} + \rho_{g})$.

\begin{figure}
  \setlength{\belowcaptionskip}{5pt}
  \centering
  \includegraphics[width=0.8\textwidth]{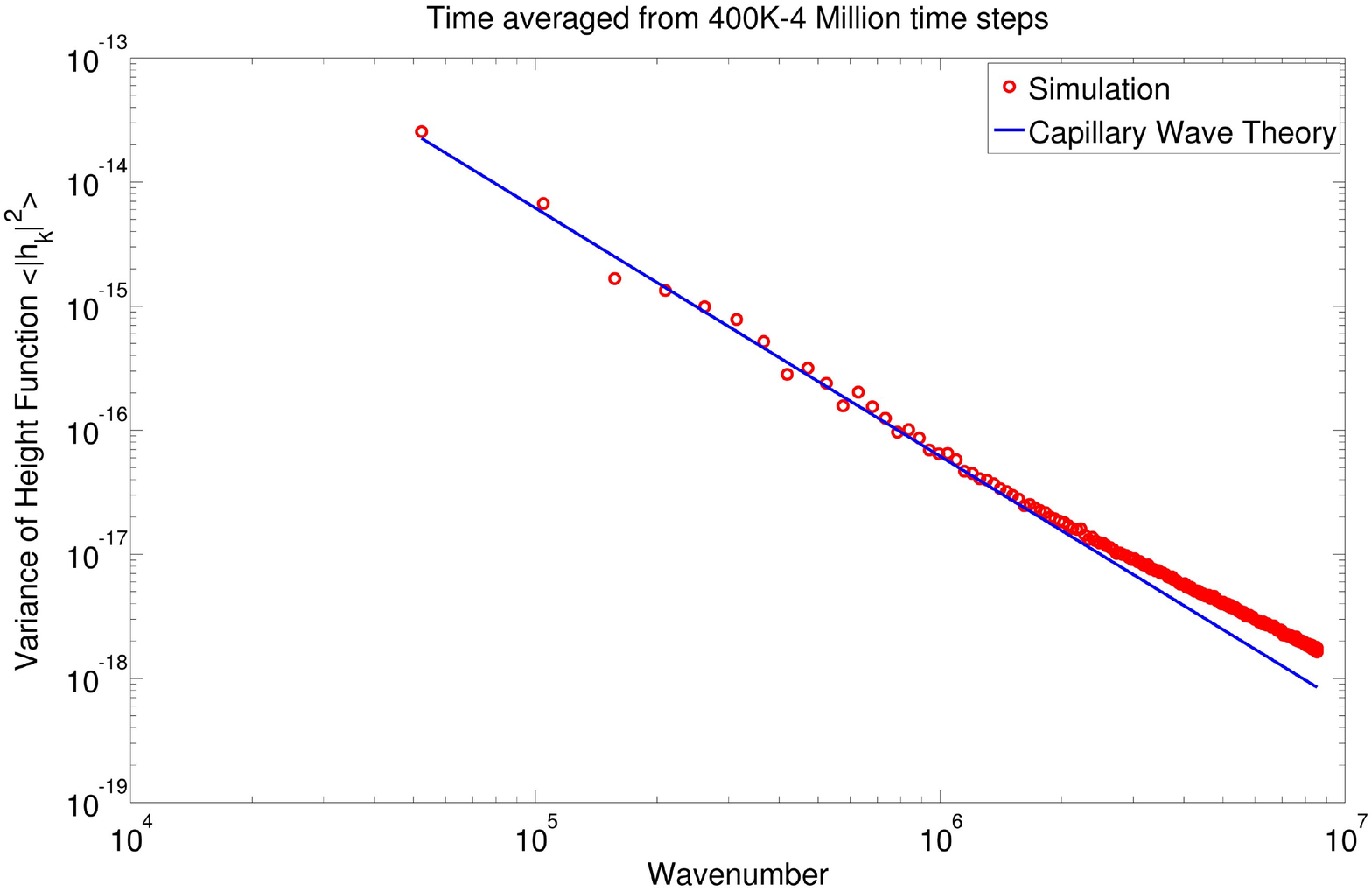}
  \caption{Variance of height fluctuations versus wavenumber comparing 2D simulations (red circles) and capillary wave theory (CWT) (black solid line); the variance is time averaged using simulation data that is saved every 1000 steps starting from 400,000 time steps and 4 million steps of total simulation time.}
  \label{FlucInterface_CWT}
\end{figure}

We simulated a $512 \times 512$ domain with $\Delta x = \Delta y = 2.34375 \times 10^{-7}~\mathrm{cm}, \Delta z = 4 \times 10^{-6}~\mathrm{cm}$ at a time step of $\Delta t = 1.25 \times 10^{-13}~\mathrm{s}$.
Simulations were initialized from the steady solution of a one-dimensional deterministic simulation that was initialized with a liquid layer 120 cells thick. The temperature was set at $T = T_{c} - 5 K$ with the liquid density set at 0.561 g/cc and vapor density at 0.256 g/cc.
The simulation was run for $4 \times 10^5$ time steps to reach statistical equilibrium before collecting data for four million time steps.
Figure \ref{FlucInterface_CWT} shows the variance of height fluctuations versus discrete wavenumber (\ref{k_discrete}) as well as the theory given by (\ref{SurfaceTension_CWT}) where the theoretical value is calculated using the discrete wavenumber for the standard Laplacian (\ref{k_discrete}).

We note that for the smaller $\Delta x$ used here, the interface is well resolved in 2D and offers more control over the noise amplitude by fine tuning the thickness of the domain; consequently, we  use $\sigma =  0.68 \; \mathrm{dynes}/\mathrm{cm}^2$
for the theory.
The data from the simulation corresponds well with the theory, indicating that the numerical method is producing results consistent with capillary wave theory. At higher wave numbers, the divergence can be attributed to compressibility effects \cite{Shang_2011}.

\section{Numerical Examples}

In this section, two computational examples are considered. First, we examine the effect of thermal fluctuations on spinodal decomposition. Next, the rapid cooling and the resulting condensation of a fluid in a square cavity with cold thermal walls is simulated.

\subsection{Spinodal Decomposition}

When a fluid is rapidly quenched from above the critical point to a thermodynamically unstable region below the critical point, the homogeneous phase separates spontaneously into coexisting phases. The domains then grow and move towards a state of minimum interfacial energy. This process of spontaneous decomposition and coarsening is called spinodal decomposition. The basic theory of spinodal decomposition was developed from a metallurgical point of view for binary alloys by Hillert \cite{Hillert_1956,Hillert_1961}, Cahn \cite{Cahn_1961,Cahn_1962}, Hilliard \cite{Hilliard_1970} and Cook \cite{Cook_1970}. The Cahn-Hilliard-Cook linear theory has been further developed and explored in a number of studies for binary alloys and binary immiscible fluids; see \cite{Langer_1975,Kawasaki_1978,Siggia_1979,Abraham_1979,Gunton_1983,Furukawa_1985,Bray_1994,Bray_2003} and references therein. Numerical studies of spinodal decomposition in liquid-vapor systems have been limited to isothermal, compressible models \cite{Nadiga_1996,Lamorgese_2009} without thermal fluctuations. In this study, we focus on a fully compressible, thermal fluid model that includes the effect of thermal fluctuations via stochastic fluxes.

\begin{figure}
  \setlength{\belowcaptionskip}{5pt}
  \vspace*{0in}
  \centering
  \includegraphics[width=.8\textwidth]{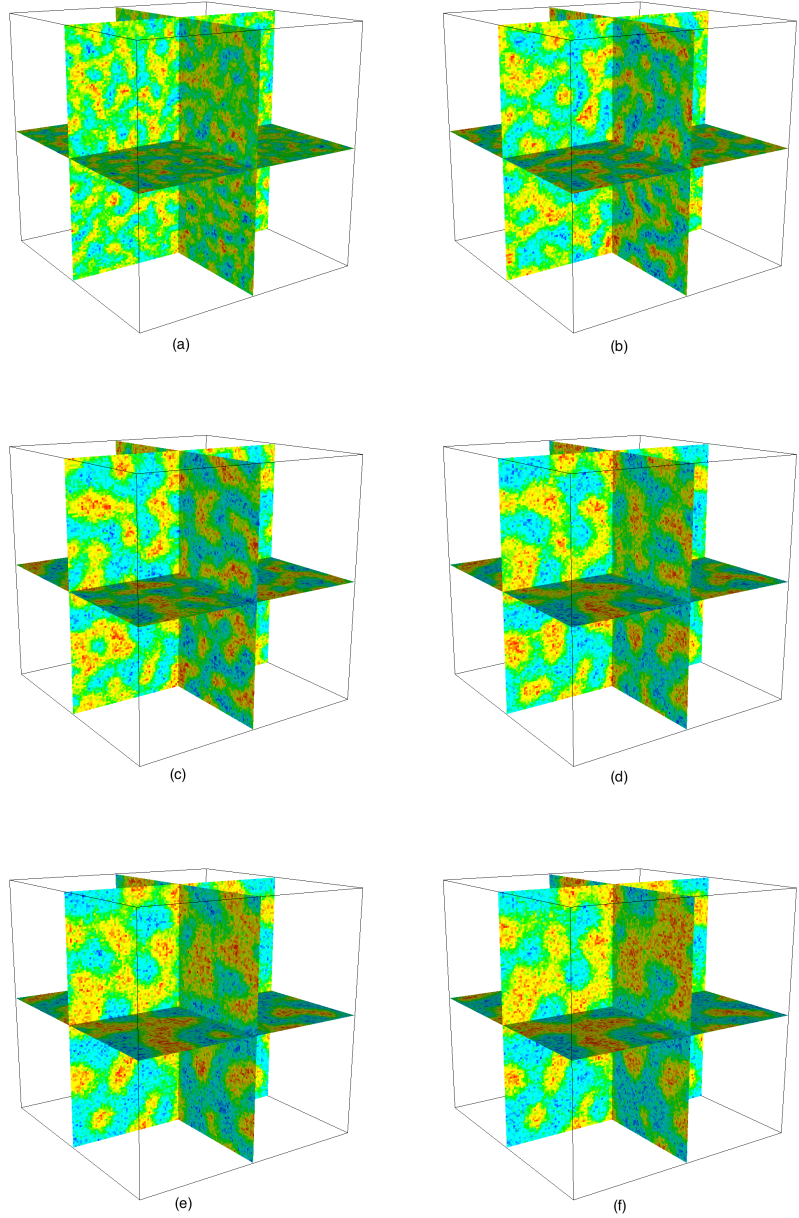}
  \caption{Liquid-vapor spinodal decomposition in a near-critical van der Waals Argon system at $\rho = 0.415995 ~\mathrm{g/cc}$, $T = 145.85 ~\mathrm{K}$ at different times: (a) $t = 2.5 \times 10^{-8}$, (b) $t = 7.5 \times 10^{-8}$, (c) $t = 1.25 \times 10^{-7}$, (d) $t = 2 \times 10^{-7}$, (e) $t = 2.5 \times 10^{-7}$ and (f) $t = 3 \times 10^{-7}$ $~\mathrm{s}$, 3D simulations with fluctuations turned on leading to formation of the bicontinuous pattern; 128\textsuperscript{3} grid points.}
  \label{3DSpinodal_Bi}
\end{figure}

The simulations were done using the parameters for an Argon fluid quenched to a temperature of $T = T_{c} - 5~\mathrm{K}$,
We first consider a 3D critical quench in which the system is initialized (see Table \ref{tab:1}) with density set to the critical density.
For this simulation, the box dimensions were $L_x = L_y = L_z = 6.4 \times 10^{-5}~\mathrm{cm}$ with a mesh size of 128\textsuperscript{3}.
The time step for this simulation was $5\times 10^{-13}~\mathrm{secs}$, again approximate 10\% of the maximum allowable
time step.   
Starting from the uniform initial state, fluctuations cause the system to decompose spontaneously into coexisting liquid and vapor regions
that form a bicontinuous pattern as shown in Fig. \ref{3DSpinodal_Bi}.
As the quenching process continues the figure illustrates the emergence on increasing larger scale features.

In order to study the role of fluctuations on the quenching process in more detail, we consider 2D systems undergoing
a critical quench, $\rho_{c} = 0.415995~\mathrm{g/cc}$, as above, and two off-critical quenches corresponding to $\rho = 0.36~\mathrm{g/cc}$ where droplets of the minority liquid phase
emerge  and $\rho = 0.47$ g/cc, which gives rise to bubbles of vapor in the majority liquid phase.
The box dimensions used in the 2D simulations were $L_x = L_y = 6 \times 10^{-5}~\mathrm{cm}$ with a thickness of $1 \times 10^{-6}~\mathrm{cm}$ and mesh size of 128\textsuperscript{2}.
The time step used is $1.0 \times 10^{-13}~\mathrm{secs}$.
Two different types of simulations were run for each quench; first where the fluctuations are present throughout the simulations and second where they are turned off after the first 5000 steps by setting the amplitudes of the stochastic fluxes to zero.
The fluctuations in the second case only serve to perturb the system from the initial state and nucleate phase separation in the quenching process.
The output data is collected after the initial $5 \times 10^{5}$ time steps up to $5 \times 10^{6}$ time steps.
In each case the process is repeated for nine different runs to compute statistics.

\clearpage

\begin{figure}
  \setlength{\belowcaptionskip}{5pt}
  \centering
  \begin{subfigure}[b]{1.0\textwidth}
    \centering
    \includegraphics[width=1.0\textwidth]{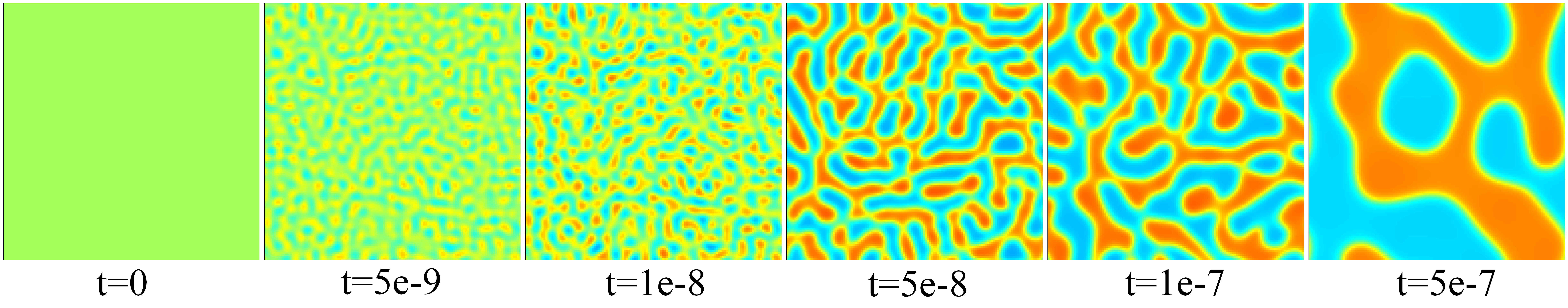}
    \caption{}
    \label{2DSpinodal_Bi_nonoise}
  \end{subfigure}
  \begin{subfigure}[b]{1.0\textwidth}
    \centering
    \includegraphics[width=1.0\textwidth]{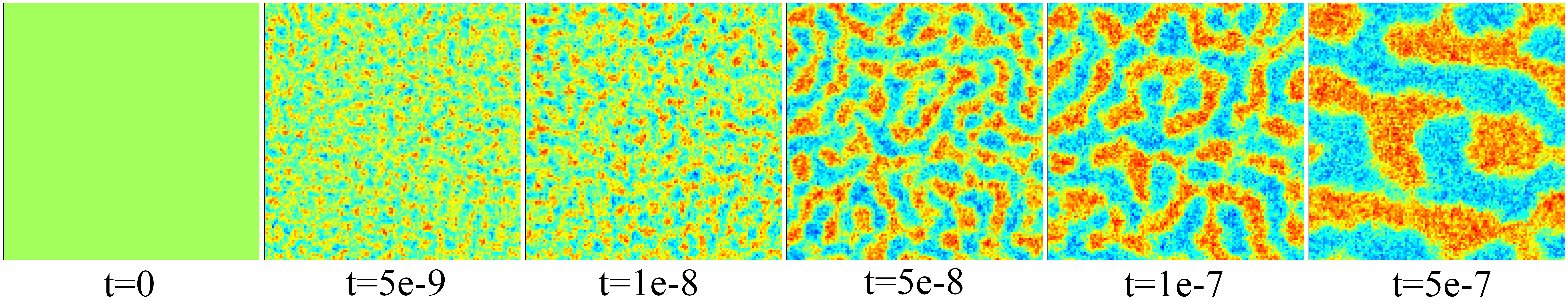}
    \caption{}
    \label{2DSpinodal_Bi_withnoise}
  \end{subfigure}
  \caption{Liquid-vapor spinodal decomposition in a near-critical van der Waals Argon system at $\rho = 0.415995 ~\mathrm{g/cc}$, $T = 145.85 ~\mathrm{K}$ at different times (secs), (a) without fluctuations and (b) with fluctuations leading to the bicontinuous pattern; 2D simulations with 128\textsuperscript{2} grid points; system volume = 36 x 10\textsuperscript{-16} cm\textsuperscript{3}.}
  \label{2DSpinodal_Bi}
\end{figure}

\begin{figure}
  \setlength{\belowcaptionskip}{5pt}
  \centering
  \begin{subfigure}[b]{1.0\textwidth}
    \centering
    \includegraphics[width=1.0\textwidth]{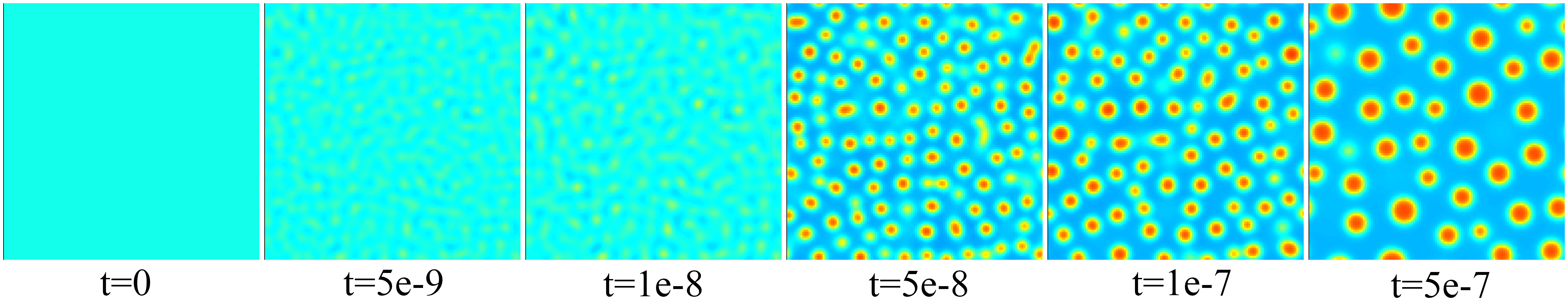}
    \caption{}
    \label{2DSpinodal_Drops_nonoise}
  \end{subfigure}
  \begin{subfigure}[b]{1.0\textwidth}
    \centering
    \includegraphics[width=1.0\textwidth]{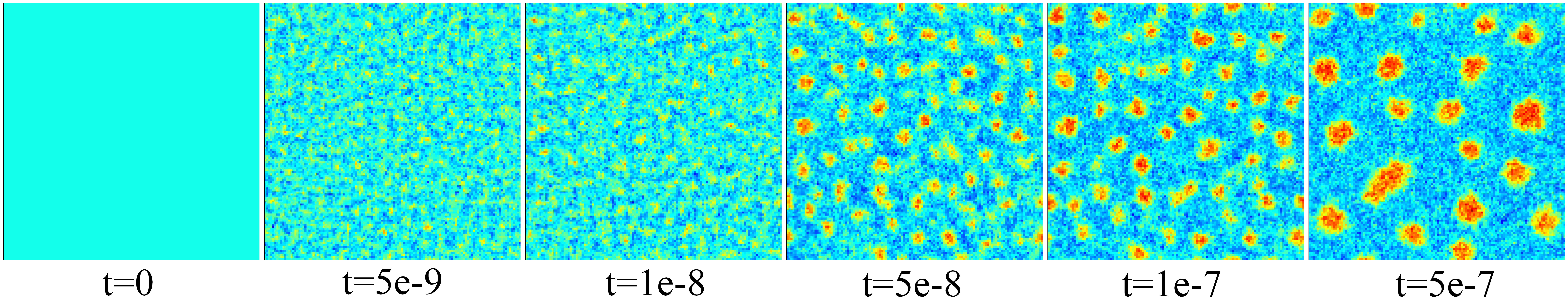}
    \caption{}
    \label{2DSpinodal_Drops_withnoise}
  \end{subfigure}
  \caption{Liquid-vapor spinodal decomposition in a near-critical van der Waals Argon system at $\rho = 0.36 ~\mathrm{g/cc}$, $T = 145.85 ~\mathrm{K}$ at different times (secs), (a) without fluctuations and (b) with fluctuations leading to the formation of droplets in a majority vapor phase; 2D simulations with 128\textsuperscript{2} grid points; system volume = 36 x 10\textsuperscript{-16} cm\textsuperscript{3}.}
  \label{2DSpinodal_Drops}
\end{figure}

\begin{figure}
  \setlength{\belowcaptionskip}{5pt}
  \centering
  \begin{subfigure}[b]{1.0\textwidth}
    \centering
    \includegraphics[width=1.0\textwidth]{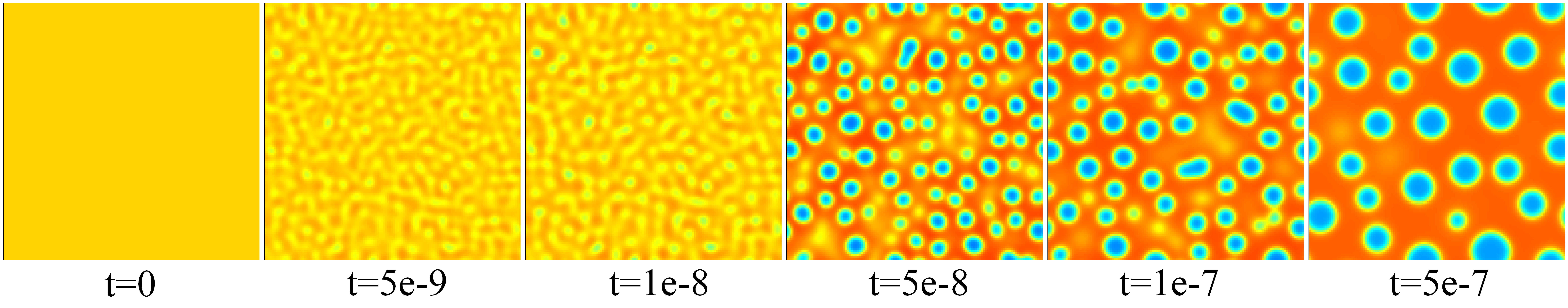}
    \caption{}
    \label{2DSpinodal_Bubbles_nonoise}
  \end{subfigure}
  \begin{subfigure}[b]{1.0\textwidth}
    \centering
    \includegraphics[width=1.0\textwidth]{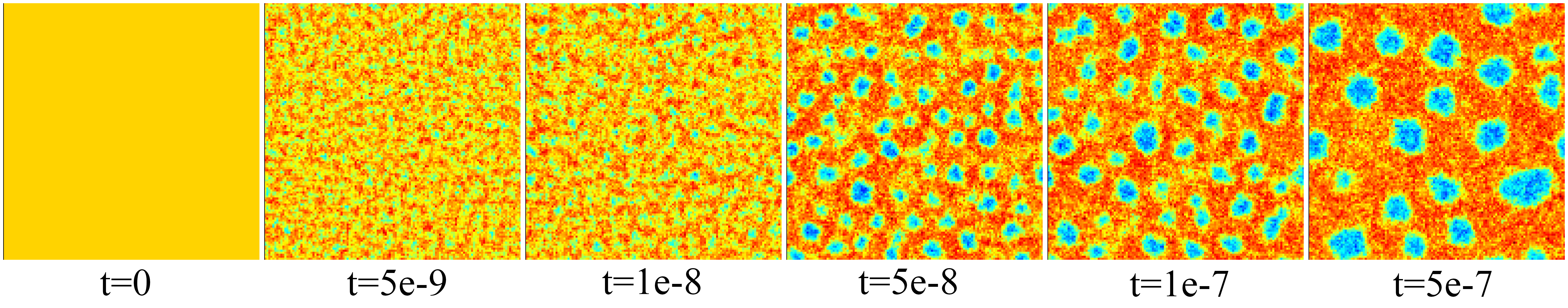}
    \caption{}
    \label{2DSpinodal_Bubbles_withnoise}
  \end{subfigure}
  \caption{Liquid-vapor spinodal decomposition in a near-critical van der Waals Argon system at $\rho = 0.47 ~\mathrm{g/cc}$, $T = 145.85 ~\mathrm{K}$ at different times (secs), (a) without fluctuations and (b) with fluctuations leading to the formation of bubbles in a majority liquid phase; 2D simulations with 128\textsuperscript{2} grid points; system volume = 36 x 10\textsuperscript{-16} cm\textsuperscript{3}.}\label{2DSpinodal_Bubbles}
\end{figure}

For the critical quench, $\rho_{c} = 0.415995~\mathrm{g/cc}$, a bicontinuous pattern similar to the 3D case emerges as shown in Fig. \ref{2DSpinodal_Bi}.
For the off-critical quench, $\rho = 0.36~\mathrm{g/cc}$, droplets of the minority liquid phase emerge as shown in Fig. \ref{2DSpinodal_Drops},
whereas the off-critical quench, $\rho = 0.47~\mathrm{g/cc}$ gives rise to bubbles of vapor in the majority liquid phase as shown in Fig. \ref{2DSpinodal_Bubbles}.
The figures also show the formation of these phases both with and without fluctuations as a function of simulation time.
It is interesting to note that Fig. \ref{2DSpinodal_Drops_withnoise} suggests that fluctuations help accelerate the growth of droplets over time.
The number of the droplets in Fig. \ref{2DSpinodal_Drops_withnoise} is larger than in Fig. \ref{2DSpinodal_Drops_nonoise} and droplets appear to be bigger in size.
In the case of the critical quench, the domains in Fig. \ref{2DSpinodal_Bi_withnoise} are less connected compared to the domains in Fig. \ref{2DSpinodal_Bi_nonoise}. The growth of bubbles in Fig. \ref{2DSpinodal_Bubbles_withnoise} follow the same trend as the droplet pattern in Fig. \ref{2DSpinodal_Drops_withnoise}.

Scaling theories of symmetric quenches in spinodal decomposition \cite{Furukawa_1985,Bray_1994} point to the existence of a single characteristic length scale in the system at late stages of decomposition when the domains have formed and are coarsening over time. During this stage, the time evolution of the system is governed by diffusive and hydrodynamic forces. For phase separation in binary fluids undergoing a critical quench, three growth regimes have been identified using scaling arguments \cite{Bray_1994},
\begin{equation}
\label{SD_ScalingLaws}
L\left(t\right) \propto  \begin{cases} t^{1/d}, & \textrm{diffusive} \\ t, & \textrm{viscous hydrodynamics} \\ t^{2/3}, & \textrm{inertial hydrodynamics} \\ \end{cases}
\end{equation}
where $d$ is the dimensionality of the system. In the diffusive regime, the system evolution is governed by the Lifshitz-Sloyozov evaporation-condensation process (Ostwald ripening) ($1/3$ power law in 3D) and Brownian coagulation ($1/3$ in 3D and $1/2$ in 2D). Hydrodynamic interactions due to interfacial effects play a significant role in the growth of domains and characterize the growth depending on the importance of the viscous and advective terms in the Navier-Stokes equation.  Furukawa \cite{Furukawa_1985} found that in 2D fluids, the viscous hydrodynamics regime is absent as the inertial forces are more important. There have been a number of numerical studies of spinodal decomposition in binary immiscible fluids and mixtures over the years to confirm some the above scaling regimes and identify others in the process \cite{SanMiguel_1985,Farrell_1989,Koga_1991,Lacasta_1992,Bastea_1995,Lookman_1996,Gurtin_1996,Gonnella_1999,Zhu_1999,Gonnella_2008,Li_2012}. Since binary fluids and liquid-vapor fall into the same universality class \cite{Hohenberg_1977}, similar scaling laws should also apply to phase separation in single component systems as well, though the precise mechanisms are not well understood in the presence of fluid motion particularly in fully compressible systems \cite{Nadiga_1996}. Previous studies on liquid-vapor isothermal models \cite{Nadiga_1996,Lamorgese_2009} have identified the growth of critical domains by a $0.70$ (\cite{Nadiga_1996}), $2/3$ (\cite{Lamorgese_2009}) power law at late stages of spinodal decomposition.

\indent There are various probes available for making a quantitative comparison of the growth rate in these systems\cite{Gunton_1983}. We chose the inverse of the weighted first moment of the structure factor, which is often used to define the characteristic length scale that grows with time,

\medskip
\begin{equation}
\label{SD_InverseMoment}
\textrm{k}_{1}^{-1} =
\dfrac{\sum_{k} \langle \delta \hat{{\rho}}(\textbf{k},t) \delta \hat{{\rho}}^*(\textbf{k},t) \rangle_k}
{\sum_{k} k \, \langle \delta \hat{{\rho}}(\textbf{k},t) \delta \hat{{\rho}}^*(\textbf{k},t) \rangle_k}
\end{equation}
\medskip

\noindent where $\langle \cdots \rangle_k$ denotes the average over a shell in Fourier space at fixed $k = \left|\textbf{k}\right|$.

\noindent The inverse of the weighted first moment is fit to a power law function of the form $A_{0} t^{A_{1}}$ to estimate the growth law exponent $A_{1}$ in each of the 9 runs for all the cases. The average growth law exponent and error (standard deviation from mean) are reported in Tables \ref{tab:4} and \ref{tab:5}.

\begin{table}[!htb]
\begin{center}
  \begin{tabular} {| c | c |}
    \hline
   Case Studied & Growth Law Exponent (Error) \\ \hline
   Droplets with Fluctuations & 0.287 (0.045) \\
   Droplets without Fluctuations & 0.208 (0.015) \\
   \hline
   Bicontinuous with Fluctuations & 0.691 (0.200) \\
   Bicontinuous without Fluctuations & 0.644 (0.191) \\
   \hline
   Bubbles with Fluctuations & 0.290 (0.053) \\
   Bubbles without Fluctuations & 0.253 (0.025) \\ \hline
  \end{tabular}
  \caption{Mean growth law exponent calculated by fitting a power law function to Eq. (\ref{SD_InverseMoment}). Mean and error (standard deviation from mean) calculated over 9 simulations runs for the entire length of saved output (4.5x10\textsuperscript{6} time steps).}
  \label{tab:4}
\end{center}
\end{table}

\begin{table}[!htb]
\begin{center}
  \begin{tabular} {| c | c |}
    \hline
   Case Studied & Growth Law Exponent (Error) \\ \hline
   Droplets with Fluctuations & 0.333 (0.039) \\
   Droplets without Fluctuations & 0.225 (0.032) \\
   \hline
   Bicontinuous with Fluctuations & 0.508 (0.076) \\
   Bicontinuous without Fluctuations & 0.511 (0.070) \\
   \hline
   Bubbles with Fluctuations & 0.289 (0.042) \\
   Bubbles without Fluctuations & 0.276 (0.073) \\ \hline
  \end{tabular}
  \caption{Mean growth law exponent calculated by fitting a power law function to Eq. (\ref{SD_InverseMoment}). Mean and error (standard deviation from mean) calculated over 9 simulations runs for the first 500,000 time steps of saved output.}
  \label{tab:5}
\end{center}
\end{table}

In Table \ref{tab:4} the power law fit is done over the entire length of the simulations output whereas it is done only over the initial $5 \times 10^5$ time steps in Table \ref{tab:5}. From the data shown in both tables, it is clear that thermal fluctuations enhance the growth of droplets over time. However, no such conclusions can be drawn for the bicontinuous or bubble cases. Density fluctuations in gases are higher than in liquids. In the droplet case, where gas is the majority phase, the enhanced density fluctuations influence the growth of droplets. In the bubble case, where liquid is the majority phase, density fluctuations are weaker and hence the effect of fluctuations is not as pronounced. It is also difficult to draw any conclusions for the bicontinuous case since there is a large overlap in the growth law exponents. The growth of critical domains without fluctuations is governed by a ~0.65 ($\pm$ 0.19) power law for the entire simulation, which is comparable to the 0.70 (\cite{Nadiga_1996}) and 2/3 (\cite{Lamorgese_2009}) power laws identified by previous studies on isothermal models.

\subsection{Cooling by the Piston Effect}

\indent In this final example we consider a homogeneous fluid, enclosed in a square cavity, at the critical density and at a temperature above the critical temperature. At $t=0$ the temperature of the cavity's walls drop below the critical temperature, initiating condensation at the walls. Due to the large compressibility of the fluid near criticality the adiabatic cooling process by sound waves is the dominant heat transfer mechanism, equilibrating the temperature far more rapidly than thermal conduction \cite{Onuki_1990}. Phase separation is induced in the fluid due to contraction of diffusive boundary layers \cite{Onuki_2007}, which is termed the piston effect ~\cite{PistonEffect1,PistonEffect2}. Due to the appearance of buoyant convection this effect cannot be reproduced experimentally under Earth's gravity, it has been observed in space under microgravity conditions \cite{MicroGravity}. This problem serves a good example of how spinodal decomposition can be induced by non-equilibrium boundary effects as opposed to isothermal situations where it is primarily studied. 

\indent In our simulations the system was initially at a supercritical state with temperature $T = T_{c} + 5 ~\mathrm{K}$ and density $\rho = \rho_{c}$. The cavity size used in 2D simulations is $\left(6\textrm{x}10^{-5}\right)^{2} ~\mathrm{cm}^{2}$ with a thickness of $1 \times 10^{-4}~\mathrm{cm}$ for low noise, $1 \times 10^{-6} \mathrm{cm}$ for high noise with a mesh size of 128\textsuperscript{2} in both cases.
The time step used is $1 \times 10^{-13} ~\mathrm{secs}$. 
Boundary conditions for pressure, temperature and velocity are set using standard conditions of no-slip walls, namely,
wall temperatures were set at $T_{\mathrm{wall}} = T_{c} - 5 ~\mathrm{K}$, wall velocities were set to zero and
pressure at the boundary condition is set so that there is no pressure gradient in the normal direction to the wall.
Because of higher-order derivative terms in the Korteweg stress an additional density boundary condition is needed.
Here we have imposed that the second normal derivative of density is zero, which makes the Korteweg stress smooth upto the boundary.

\begin{figure}
  \setlength{\belowcaptionskip}{0pt}
  \vspace*{0in}
  \centering
  \includegraphics[width=1\textwidth]{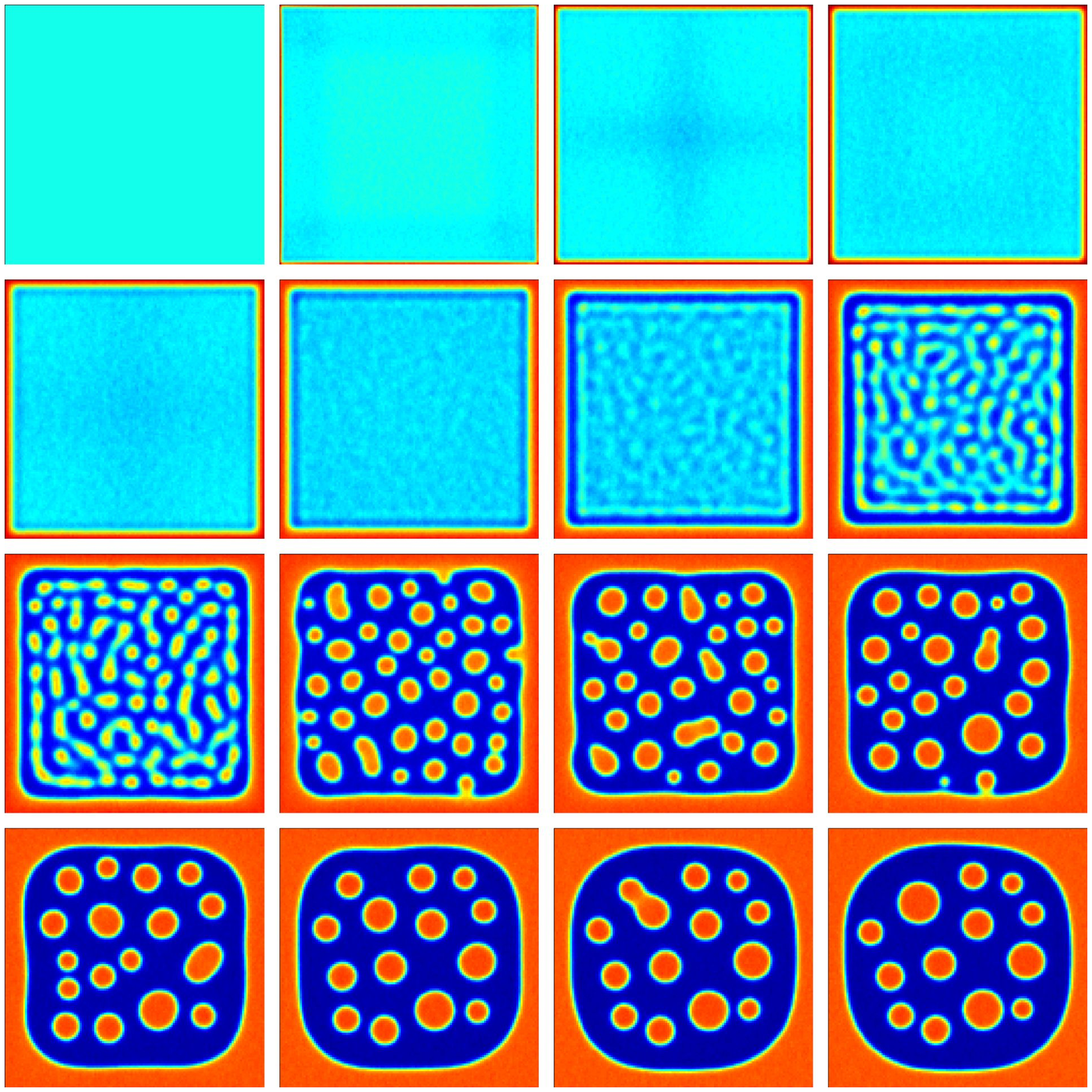}
  \caption{Adiabatic expansion of super-critical Argon in a square cavity driven by boundary cooling at different times (secs); low noise (cell volume = 2.197266 x 10\textsuperscript{-17} cm\textsuperscript{3}), left to right, top to bottom : $t = 0, t = 5 \times 10^{-10}, t = 1.5 \times 10^{-9}, t = 2.5 \times 10^{-9}, t = 4.5 \times 10^{-9}, t = 9 \times 10^{-9}, t = 1.8 \times 10^{-8}, t = 3.15 \times 10^{-8}, t = 4.5 \times 10^{-8}, t = 9 \times 10^{-8}, t = 1.26 \times 10^{-7}, t = 1.845 \times 10^{-7}, t = 2.52 \times 10^{-7}, t = 3.51 \times 10^{-7}, t = 4.545 \times 10^{-7}, t = 5 \times 10^{-7} \mathrm{s}$; the plots follow the initial development of a liquid boundary layer in the cavity due to thermalization by fast acoustic waves, followed by spinodal decomposition in the interior and thickening of both boundary layers and interior domains over time.}
  \label{AdExp_lownoise}
\end{figure}

\begin{figure}
  \setlength{\belowcaptionskip}{0pt}
  \vspace*{0in}
  \centering
  \includegraphics[width=1\textwidth]{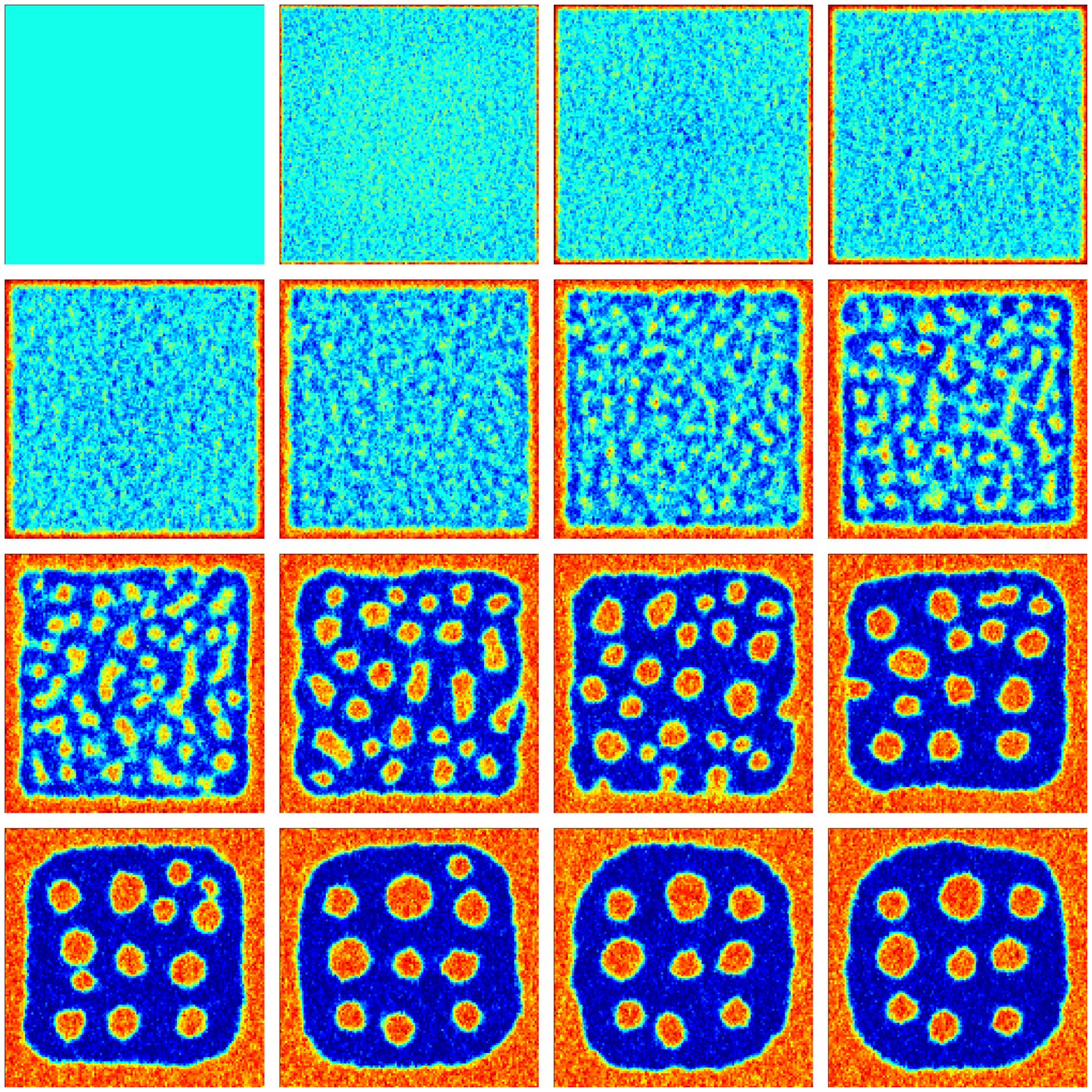}
  \caption{Adiabatic expansion of super-critical Argon in a square cavity driven by boundary cooling at different times (secs); high noise (cell volume = 2.197266 x 10\textsuperscript{-19} cm\textsuperscript{3}); simulated times same as Fig. \ref{AdExp_lownoise}; the plots follow the initial development of a liquid boundary layer in the cavity due to thermalization by fast acoustic waves, followed by spinodal decomposition in the interior and thickening of both boundary layers and interior domains over time; higher fluctuations lead to faster growth of domains and fewer droplets compared to Fig. \ref{AdExp_lownoise}.}
  \label{AdExp_highnoise}
\end{figure}

\begin{figure}
  \setlength{\belowcaptionskip}{5pt}
  \vspace*{0in}
  \centering
  \includegraphics[width=.8\textwidth]{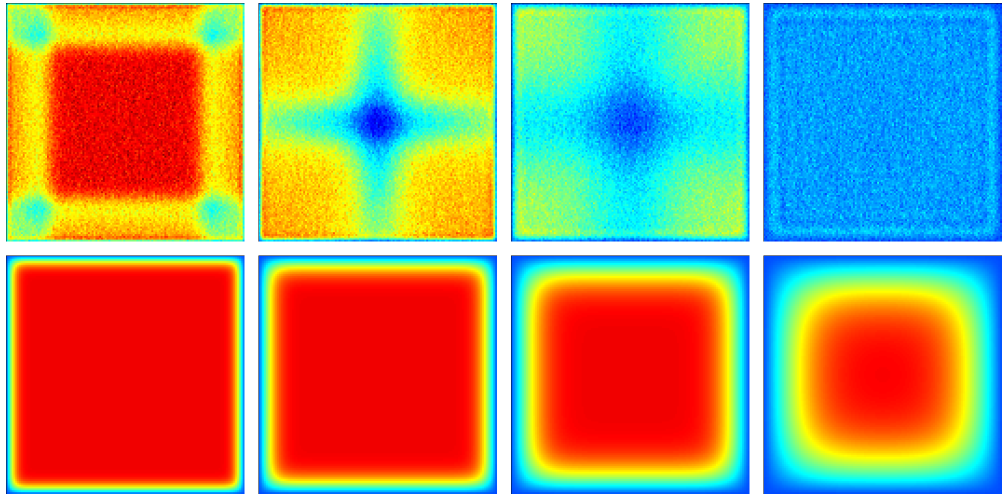}
  \caption{Temperature profile for \textit{Top Panel}: cooling by adiabatic expansion of super-critical Argon in a square cavity driven by boundary cooling, and \textit{Bottom Panel}: for cooling by heat conduction alone. Both systems start at same initial condition; plots are snapshots of temperature profile (left to right) at $t = 5 \times 10^{-10} \mathrm{s}$, $t = 1.5 \times 10^{-9} \mathrm{s}$, $t = 4.5 \times 10^{-9} \mathrm{s}$, $t = 1.25 \times 10^{-8} \mathrm{s}$; low noise (system volume = $36 \times 10^{-14} \mathrm{cm}^3$)}
  \label{AdExp_Temp}
\end{figure}

\indent When the boundary temperature is changed, pressure perturbations from acoustic waves travel and change rapidly across the domain as seen in Figs. \ref{AdExp_lownoise} \& \ref{AdExp_highnoise} for low noise and high noise respectively (large and small volume respectively; since the simulations are quasi-2D, the noise amplitude is controlled by changing the thickness of the cell). Thermal equilibration is reached through the boundary diffusive layers that act as a piston adiabatically changing the density in the cavity \cite{Onuki_1990,Onuki_2007}. To see if the temperature equilibration is indeed faster in the piston effect, the adiabatic cooling problem is compared with a simple diffusive cooling (heat equation) and the temperature profiles are compared in Fig. \ref{AdExp_Temp}. It can be seen from the figures that cooling due to piston effect causes temperatures to equalize all over the domain in the first few thousand time steps compared to diffusive cooling by Fourier's Law where the temperature profile is still roughly parabolic. The rapid density and temperature changes are followed by slow nucleation and coarsening from the boundary to the interior of the domain (see Figs. \ref{AdExp_lownoise} \& \ref{AdExp_highnoise}). The results here show qualitative agreement with deterministic simulations of Onuki \cite{Onuki_2007}.
When the effect of thermal fluctuations is increased, the pressure perturbations and coarsening happen on a faster time scale as seen in Fig. \ref{AdExp_highnoise}. 

\section{Conclusions}

Fluctuating hydrodynamics is a powerful mesoscopic technique that combines fluid mechanics and statistical mechanics to model physical systems more realistically. The compressible fluctuating hydrodynamics equations bridge the gap between molecular and hydrodynamic length and time scales. Numerical methods for solving these stochastic PDEs for single species or multi-species fluids are now well-established. Yet many important applications are both multi-component and multi-phase fluids and it is important to include all the relevant physics when modeling these complex systems.
As a first step in this direction we have extended the existing fluctuating hydrodynamics framework to model two-phase fluids.

A van der Waals diffuse interface model was developed in this study for solving the fluctuating Landau Lifshitz Navier Stokes equations. Our finite volume scheme treats the advective, diffusive, Korteweg, and stochastic fluxes consistently and uses an accurate stochastic, three-stage Runge-Kutta temporal integrator. The appropriate value for the gradient energy coefficient, $\kappa$, was established based on numerical tests done to reproduce the correct surface tension of Argon near the critical point. The scheme was validated by calculating the 2D and 3D structure factors of density and comparing it with analytical expressions. Additional validation was done by comparing the capillary wave spectrum obtained from simulations with the theoretical prediction.

Two non-equilibrium systems were simulated to demonstrate the utility of the scheme. In the first example, spinodal decomposition in a near-critical van der Waals Argon fluid, the scheme captures the essential physics quite well by reproducing the expected bicontinuous pattern for critical quenches and droplets/bubbles for off-critical quenches. Calculations of the growth law using the inverse first moment of the structure factor compare well with existing studies in the literature. They point additionally to the idea that fluctuations enhance droplet growth. The second example examined droplet formation in a square cavity due to adiabatic cooling by sound waves (piston effect)initiated when the boundary temperature is lowered suddenly. The simulations agree qualitatively with earlier deterministic calculations but also indicate that thermal fluctuations lead to a faster growth of droplets.

The multiphase fluctuating hydrodynamics methodology was developed in this paper for near-critical Argon with constant transport properties. This can be extended for fluids very close to the critical point where the transport properties are now wavenumber dependent, especially thermal diffusivity \cite{Sengers_IUPAC}. If a system is very close to the critical point, the thermodynamic and transport properties are governed by asymptotic critical behavior, whereas far away from the critical point, the behavior is more classical thermodynamically. The methodology can be extended to systems far away from the critical point using cross-over equations of state \cite{Anisimov1992crossover} that help connect the critical phenomena with classical thermodynamics. There is also a need to go beyond the van der Waals theory \cite{van1985gravity, Sikkenk1986gravity, Sengers1989capillary, van1989capillary} to include gravitational effects and study the effect of fluctuations on the interface and capillary wave spectrum.

A possible extension of this work would be for systems with non-constant $\kappa$; i.e., $\kappa(\rho,T)$. Dependencies of this kind would however lead to additional terms that can arise in the momentum and energy equations that have to be properly admitted using the GENERIC framework. Laying the foundation for addition of cross-coupling terms such as the one we use for $\KortewegCross$, is another important avenue for future work. A number of potential applications in the area of complex fluids lie in the multiphase, multispecies (multicomponent) mixtures. A natural extension of this work would be to connect to the fluctuating hydrodynamics of multispecies mixtures \cite{Balakrishnan2014}. Multi-species mixtures introduce a lot of complicated cross-diffusion effects \cite{Espanol2003} that can complicate the problem enormously. Further extension to problems with complex chemical reactions is another possibility. A large number of interesting applications involve drops or bubbles either suspended/moving in liquids or on a solid surface. The framework developed here could be extended to model bubble dynamics in biomedical applications such as drug delivery \cite{Lohse_2003}, gas embolism dynamics where a surface tension gradient due to surfactant adsorption at the interface can lead to Marangoni stresses that cause bubble motion \cite{Swaminathan_2010}, or in dynamical wetting transitions at large driving velocities to study contact line motion \cite{Snoeijer_2013}. Fluctuations are also extremely important in applications where different types of particles/ions are driven either towards or away from the interface \cite{Noah_2009}.

One of the challenges that still remains is the time step limitation due to the explicit nature of the integration algorithm. The acoustic modes limit the time steps that can be taken in the system thus causing the simulations to run slower and longer. One way to overcome this is to look for low Mach number formulations where sound modes are eliminated from the compressible equations. While acoustic waves are important at short times as in the piston problem, low Mach number schemes will be important for the long time limit. Analysis of such low Mach formulations for phase change fluctuating hydrodynamics is very challenging and will be the subject of future research.

Another numerical challenge is the competing requirements of having a fine mesh to completely resolve the interface versus the size of the cell volume that is directly proportional to the amplitude of the fluctuations. The numerical algorithms have to be designed very carefully for both of the above requirements to be met. On way of doing this by filtering the stochastic fluxes has already been discussed in \cite{LowMachExplicit}. However, more work needs to be done to extend this to multiphase, multispecies fluctuating hydrodynamics.

\section*{Acknowledgements}

The work at LBNL was supported by the Applied Mathematics Program of the U.S. DOE Office
of Advance Scientific Computing Research under contract DE-AC02005CH11231.
A. Donev was supported in part by the National Science Foundation under grant DMS-1115341
and the Office of Science of the U.S. Department of Energy through Early Career award number DE-SC0008271.
A.C. would also like to thank Dr. Andy Nonaka at Lawrence Berkeley Lab for his help with creating the plots for 3D structure factors.

\pagebreak
\bibliographystyle{unsrt}
\bibliography{FlucvdWLLNS_refs}
\end{document}